\documentclass[sigconf]{acmart}
\usepackage{array}
\usepackage{longtable}
\usepackage{ragged2e}
\usepackage{soul}
\usepackage{multirow}
\usepackage{subcaption}
\usepackage{makecell}
\usepackage{tabularx}
\usepackage{placeins}
\usepackage{booktabs}

\newcolumntype{P}[1]{>{\RaggedRight\arraybackslash}p{#1}}

\AtBeginDocument{%
  }

\setcopyright{acmlicensed}
\copyrightyear{2026}
\acmYear{2026}
\setcopyright{cc}
\setcctype{by}
\acmConference[CHI '26]{Proceedings of the 2026 CHI Conference on Human Factors in Computing Systems}{April 13--17, 2026}{Barcelona, Spain}
\acmBooktitle{Proceedings of the 2026 CHI Conference on Human Factors in Computing Systems (CHI '26), April 13--17, 2026, Barcelona, Spain}
\acmDOI{10.1145/3772318.3790966}
\acmISBN{979-8-4007-2278-3/2026/04}

\usepackage{multirow}

\usepackage{float}
\begin{document}

\title{PASTA: A Scalable Framework for Multi-Policy AI Compliance Evaluation}

\author{Yu Yang}
\email{yyang124@student.ubc.ca}
\affiliation{%
  \institution{The University of British Columbia}
  \city{Vancouver}
  \country{Canada}
}

\author{Ig-Jae Kim}
\email{drjay@kist.re.kr}
\affiliation{%
  \institution{Korea Institute of Science and Technology}
  \city{Seoul}
  \country{South Korea}
}

\author{Dongwook Yoon}
\email{yoon@cs.ubc.ca}
\affiliation{%
  \institution{The University of British Columbia}
  \city{Vancouver}
  \country{Canada}
}

\begin{abstract}
AI compliance is becoming increasingly critical as AI systems grow more powerful and pervasive. Yet the rapid expansion of AI policies creates substantial burdens for resource-constrained practitioners lacking policy expertise. Existing approaches typically address one policy at a time, making multi-policy compliance costly. We present PASTA, a scalable compliance tool integrating four innovations: (1) a comprehensive model-card format supporting descriptive inputs across development stages; (2) a policy normalization scheme; (3) an efficient LLM-powered pairwise evaluation engine with cost-saving strategies; and (4) an interface delivering interpretable evaluations via compliance heatmaps and actionable recommendations. Expert evaluation shows PASTA's judgments closely align with human experts ($\rho \geq .626$). The system evaluates five major policies in under two minutes at approximately \$3. A user study (N = 12) confirms practitioners found outputs easy-to-understand and actionable, introducing a novel framework for scalable automated AI governance.
\end{abstract}



\maketitle

\section{Introduction}
    As AI technologies continue to grow in power and evolve at an unprecedented pace ~\cite{TANG2020101094, HAJKOWICZ2023102260, oecd2023, stanford2025}, ensuring policy compliance becomes increasingly critical for supporting the Responsible AI (RAI) lifecycle. This need arises in part from the substantial risks introduced by AI, as the rapid pace of innovation further increases its influence in almost every aspect of people's day-to-day lives ~\cite{hendrycks2023overviewcatastrophicairisks}. In addition, the scale and autonomy of modern AI systems have contributed to a decline in public trust, driven by concerns about insufficient transparency and oversight ~\cite{10.1145/3442188.3445890, kierans2025catastrophicliabilitymanagingsystemic}. Ensuring compliance with policy frameworks is therefore essential not only for mitigating these risks but also for promoting explainability and traceability, which are foundational to building a trustworthy and sustainable AI landscape.

    The landscape of AI governance is growing in complexity as AI practitioners face a flood of rapidly evolving policies. By May 2023, more than 930 AI-related policy initiatives had been introduced in 71 jurisdictions around the world ~\cite{oecd2023}. This momentum has only been accelerating, with AI-related legislative activity growing by 21.3\% in 75 countries between 2023 and 2024 and the number of AI-related federal regulations in the United States doubling within a year ~\cite{stanford2025}. These policies vary widely in scope and jurisdiction, targeting different aspects of AI lifecycles and geographical applications. As a result, modern AI systems must undergo \emph{multi-policy, cross-jurisdictional} compliance checks to ensure they satisfy these diverse and continually shifting regulatory requirements ~\cite{nannini2023explainabilityaipoliciescritical}. In this paper, we define the ability to effectively manage compliance across diverse policies as scalability. 

    Executing compliance evaluation purely by hand does not scale. Traditionally, many organizations have relied on hiring specialized legal and compliance teams to manually interpret policies and ensure regulatory alignment ~\cite{iapp2025aigov}. The workload grows exponentially as each system capability must be assessed against numerous provisions. Beyond time and cost, manual processes struggle with traceability (linking findings to evidence) and comparability across assessors. To make routine, repeatable, and auditable evaluations possible, especially for teams without dedicated legal staff, practitioners need tool-supported workflows to maintain compliance at scale.

    To support compliance evaluation across a diverse policy landscape, compliance tools must address design challenges in their input and output mechanisms. On the input side, scalable system–policy comparison requires a standardized format that can consistently map system descriptions to heterogeneous policy requirements. The challenge in designing such a format is balancing the comprehensiveness needed to capture system details for multi-policy compliance with the simplicity required for practitioners to complete it with reasonable effort ~\cite{aca5311e38a64463a612bbff5f4198ca}. On the output side, scalability poses interpretability challenges, as extensive policy coverage produces extensive results. Without a clear and structured representation, these evaluations risk becoming too dense or fragmented for practitioners to understand and impose substantial sensemaking burdens from an HCI perspective. This issue is also compounded by the complexity of AI policies themselves, which are often lengthy, technical, and written in legal language that is difficult to navigate without specialized expertise. To ensure that large-scale evaluations remain actionable, compliance tools must organize complex results into understandable formats to help users identify key issues and make informed decisions.

    Even with scalable, user-friendly designs, cost remains a major barrier to adopting AI compliance evaluation methods. Large language models (LLMs) provide a promising alternative by enabling automation in compliance evaluation ~\cite{10628489, wang2025llmbasedhsecomplianceassessment, 10628503}. However, LLMs come with their own computational and financial costs, especially when processing long policy documents. For example, the California Consumer Privacy Act (CCPA), the European Union Artificial Intelligence Act (EU AI Act), and the Artificial Intelligence and Data Act (AIDA) contain an average of 66 articles, with approximately 374 words per article ~\cite{ccpa2018, eu2024_regulation1689, billc27_2022}. When combined, the total text across these three policies exceeds 74,000 words. Longer inputs and outputs make LLM requests more costly and slower, as they increase both token usage and processing time ~\cite{OpenAI2025Pricing}. As a result, evaluating compliance across multiple policies often involves large-scale, multi-step comparisons that can lead to considerable computational and financial burdens. These cost considerations are especially critical as the number of applicable policies continues to grow, demanding solutions that balance scalability with cost and time efficiency in LLM-powered compliance evaluation.

    Current research prototypes for AI policy compliance evaluation are often tailored to a single major policy ~\cite{herdel2024exploregenlargelanguagemodels, Agora2023} or rely on complex technical input requirements ~\cite{Farsight2024, 10.1145/3685651.3686700, madaio2024tinker}. Similarly, commercial solutions typically approach AI governance using: (1) policy-specific questionnaires (e.g., LumeNova, VerifyWise, Collibra AI, Credo AI), which hard-code regulations into survey logic and require one survey per policy, offering no flexible or unified input format ~\cite{LumenovaAI, VerifyWise, CollibraAI2024, credo2025platform}; and (2) code-scanning or automated workflows (e.g., OneTrust, OpenLayer, Vanta), which depend on access to system metadata or source code and therefore only support compliance once a system is already implemented ~\cite{OneTrustWebsite, Openlayer2025, VantaWebsite}. Their high costs also make them inaccessible to individual practitioners and smaller teams. These approaches leave gaps in flexibility, early-stage usability, and scalability. To address these limitations, we introduce PASTA (Policy Aggregator \& Scanner for Trustworthy AI), an AI compliance evaluation tool that leverages LLMs to assess AI system documentation across multiple global policies, including the EU AI Act, AIDA, GDPR, CCPA, and the Colorado AI Act. PASTA is designed for individual practitioners (e.g., developers, project managers, and UI designers) or small teams who typically (1) lack expertise in AI policy compliance and legal knowledge, and (2) operate with limited resources. PASTA provides a multi-policy framework with relatively low-effort input design and interpretable compliance outputs, making AI policy evaluation more accessible and actionable. Specifically, our system addresses the existing gap in the compliance evaluation tool landscape by:
    
\begin{itemize}
    \item To enable scalability in our policy database, PASTA incorporates a policy-structuring framework that unifies diverse policy formats into a consistent, structured, paragraph-level table format.
    
    \item To power scalable compliance evaluation across policies, PASTA is designed around a streamlined model-card input mechanism and multi-dimensional output interface that enable scalable and interpretable assessments.
    
    \item To enhance cost efficiency, PASTA employs Policy Chunking, which partitions policies into smaller, coherent units, and Irrelevancy Mapping, which filters out non-informative comparisons. Together, these strategies reduce the number of LLM queries, lowering costs and latency while maintaining comprehensive policy coverage.
\end{itemize}
     
    To assess the usability and accuracy of PASTA, we conducted both a two-part user study with 12 AI practitioners and a structured expert-based technical evaluation. Results showed that PASTA enabled comprehensive multi--policy assessments while keeping the entry barrier manageable: participants completed the model card input in an average of 28.4 minutes ($\mathrm{SD} = 6.7$) and described the process as cognitively effortful yet highly valuable for reflecting on system design and documentation. The resulting compliance reports were viewed as actionable and interpretable, with users confidently identifying policy-specific risks ($\mathrm{M} = 4.73$, $\mathrm{SD} = 0.45$) and linking them to system components in under seven minutes on average. Our technical evaluation demonstrated that PASTA’s violation and relevance scores closely aligned with expert consensus (Spearman $\rho = 0.6264$ and $\rho = 0.7611$), with mean absolute errors remaining low and 87--94\% of predictions falling within one point of expert labels. Taken together, these findings indicate that PASTA not only supports scalable and interpretable compliance evaluation but also produces outputs that are directionally consistent with legal expert reasoning, shaping practitioners’ understanding of AI compliance and prompting earlier integration of compliance considerations in their workflows.

    In this work, we present PASTA, the first automated and scalable solution for multi-policy AI governance. PASTA introduces (1) a model-card–based input schema that balances ease of use with completeness; (2) a policy normalization pipeline that restructures diverse regulations into a consistent format for LLM evaluation; (3) cost-saving algorithmic strategies, including dynamic chunking and irrelevancy mapping; and (4) an interpretable reporting interface validated through a practitioner study to support actionable, multi-policy compliance insights.

\section{Related Work}
As introduced in Section 1, the current AI compliance landscape faces challenges of scalability and cost-efficiency. In this section, we review existing approaches in more depth and position this work within the broader research landscape.

\subsection{Definition of AI Governance and AI Compliance}

PASTA is specifically designed to address the challenges of AI compliance. To situate our work, we first provide essential background on the broader landscape of AI governance and its relationship to AI compliance.

AI governance provides the broader structures that guide the development, deployment, and oversight of artificial intelligence systems so they align with legal, ethical, and societal expectations ~\cite{Arnold_Schiff_Schiff_Love_Melot_Singh_Jenkins_Lin_Pilz_Enweareazu_Girard_2024, mucci2024_ibmai_gov, dafoe2018ai}. It encompasses governance instruments, enforcement mechanisms, and the implementation of responsible AI principles throughout the system lifecycle, and has been described by Floridi and Cowls as a ``normative framework of policies, principles, tools, and institutions'' aimed at safeguarding fundamental rights and democratic values ~\cite{Floridi2019Unified}. Within this broader landscape, AI compliance operates as a focused domain concerned with verifying conformance to legal, regulatory, and ethical standards ~\cite{BARREDOARRIETA202082, 10616752}. In line with the EU AI Act, which defines an AI system as “a machine-based system… designed to operate with varying levels of autonomy” and capable of generating outputs that influence physical or virtual environments ~\cite{eu2024_regulation1689}, we adopt a broad definition of compliance that spans both legal and ethical dimensions. Likewise, we use the term policy to refer to both binding regulations and non-binding guidelines, recognizing that compliance work must integrate these diverse sources of obligation ~\cite{schöning2025complianceaisystems}.

\subsection{Existing Landscape of Automated Tools within AI Governance (Algorithmic Auditing)}
A growing body of research has emerged to address the challenges of ensuring compliance in AI systems. Building on a recent review synthesizing the breadth of research supporting AI governance ~\cite{batool2023responsibleaigovernancesystematic, 10.1145/3626234}, we group existing work into two main approaches: conceptual methodologies and automated tools. Conceptual methodologies include governance frameworks, ethical principles, policies, and high-level guidelines that offer theoretical direction but often lack enforcement mechanisms ~\cite{raji2020closingaiaccountabilitygap, 10.1145/3613904.3642849}. In contrast, automated tools instead translate requirements into executable checks and algorithmic auditing. Our work situates within automated AI governance tools, emphasizing enhancing accessibility and scalability for AI practitioners without legal expertise or support. To structure our review, we further classify existing automated tools into four categories: Risk Prediction Tools, System Behavior Verification Tools, Policy Interpreters, and Commercial AI Governance Platforms.

\paragraph{Risk Prediction Tools}
A class of automated AI governance tools supports early-stage development by simulating use cases and predicting potential harms using large language models (LLMs). Some tools, such as ExploreGen and Farsight, generate plausible use cases, stakeholders, and associated risks directly from system descriptions to help practitioners surface overlooked scenarios ~\cite{herdel2024exploregenlargelanguagemodels, Farsight2024}. Others provide more structured guidance: the AI Risk Atlas consolidates known generative- and agentic-AI risks and links them to benchmarks and mitigation strategies ~\cite{bagehorn2025airiskatlastaxonomy}, while Usage Governance Advisor derives risk assessments and safeguards from a user’s described intent ~\cite{daly2025usagegovernanceadvisorintent}. These tools embed early risk discovery into prototyping workflows, but they focus narrowly on use-case harms and typically overlook broader compliance needs such as data governance, transparency documentation, accountability mechanisms, fairness evaluation, and alignment with legal requirements.

\paragraph{System Behavior Verification Tools}
Another stream of AI governance tools focuses on system accountability by monitoring model behavior and verifying decision-making processes ~\cite{varshney2018introducing, 10.1145/3685651.3686700, 10.1145/3706598.3713301, Lam_2025}. IBM’s Fairness Tool offers real-time analysis to detect biases in algorithmic outputs ~\cite{varshney2018introducing}, while Cyberismo evaluates whether software systems satisfy cybersecurity requirements from user-provided descriptions ~\cite{10.1145/3685651.3686700}. Although these tools improve transparency, fairness, and code security tracking, their reliance on manual input and fixed logic rules makes them difficult to adapt to evolving governance needs, diverse architectures, and cases where source code cannot be shared or has not been implemented ~\cite{10.1145/3706598.3713301}.

\paragraph{Policy Interpreters}
Efforts to summarize, transform, and make complex legal texts actionable have led to AI governance tools designed to help practitioners more effectively navigate compliance requirements ~\cite{10.1145/3675888.3676142, Arnold_Schiff_Schiff_Love_Melot_Singh_Jenkins_Lin_Pilz_Enweareazu_Girard_2024, constantinides2024raiguidelinesmethodgenerating}. Many use LLMs and human-centered methods to interpret regulations and generate concise summaries. For example, Mittal et al. automate GDPR privacy-policy interpretation to extract PII-related requirements and produce compliance-oriented summaries ~\cite{10.1145/3675888.3676142}. Meanwhile, Arnold et al.’s AGORA dataset offers a structured, cross-jurisdictional taxonomy of AI laws to support regulatory comparison ~\cite{Arnold_Schiff_Schiff_Love_Melot_Singh_Jenkins_Lin_Pilz_Enweareazu_Girard_2024}. Although such tools make regulatory information more accessible, they focus primarily on summarization and provide limited compliance evaluation features.

\paragraph{Commercial AI Governance Platforms.}

\begin{table*}[t]
\centering
\caption{Key functional capabilities, regulatory coverage, and pricing across representative AI governance tools. The ``Core Capabilities'' column aggregates features where the vendor explicitly provides support.}
\label{tab:governance-tools}

\footnotesize 

\begin{tabular}{p{2cm} p{5cm} p{4cm} p{3.2cm}}
\toprule
\textbf{Tool} &
\textbf{Core Capabilities} &
\textbf{Supported Policies} &
\textbf{Pricing} \\
\midrule

Atlan &
System Registry; Risk Assessment; Model Testing &
None &
Starting from \$100{,}000/year \\
\midrule

Credo~AI &
System Registry; Risk Assessment; Compliance Assessment &
Colorado SB~21-169; EU AI Act Readiness; ISO~42001; NIST AI RMF; NYC Local Law~144 &
Not publicly available \\
\midrule

IBM Watsonx.governance &
System Registry; Risk Assessment; Model Testing &
None &
\$36{,}000/year (1 instance, 5 use cases, 25 users, 12k evaluations) \\
\midrule

Collibra AI Governance &
System Registry; Risk Assessment; Model Testing; Compliance Assessment &
EU AI Act; NIST AI RMF &
\$170{,}000/year \\
\midrule

OneTrust &
System Registry; Risk Assessment; Model Testing; Compliance Assessment &
EU AI Act; ISO~42001; NIST RMF &
\$827--\$2{,}275/month (Based on policy) \\
\midrule

VerifyWise &
System Registry; Risk Assessment; Model Testing; Compliance Assessment &
EU AI Act; ISO~42001 &
\$799/month (5 use cases + 2 frameworks) \\
\midrule

LumeNova &
System Registry; Risk Assessment; Model Testing; Compliance Assessment &
Colorado Reg.~10-1-1; EU AI Act; ISO~42001; NIST AI RMF; NYC Local Law~144 &
Not publicly available \\
\midrule

Vanta &
System Registry; Risk Assessment; Compliance Assessment &
35+ frameworks (e.g., ISO~27001, SOC~2, HIPAA) &
Starting at \$14{,}000/year \\
\midrule

OpenLayer &
System Registry; Risk Assessment; Model Testing; Compliance Assessment &
ISO/IEC~42001; OWASP; NIST; EU AI Act &
Not publicly available \\
\midrule

Amazon Bedrock (Guardrails) &
System Registry; Model Testing &
None &
\$29{,}462.40/month for 1 Claude Instant unit running 24/7 in US West (Oregon) Region \\
\bottomrule
\end{tabular}
\end{table*}

A broad ecosystem of commercial AI governance platforms has emerged to support organizations in managing AI systems across their lifecycle. These platforms are designed primarily for enterprise regulatory workflows, where teams coordinate model registration, risk assessment, testing and documentation ~\cite{credo2025platform, CollibraAI2024, OneTrustWebsite, VerifyWise, LumenovaAI, VantaWebsite, Openlayer2025, aws2025guardrails, atlan2024, ibm2025watsonx}. Our analysis of 10 prominent platforms shows that, despite differences in emphasis, they converge around a common set of functional components:

\begin{itemize}
    \item \textbf{AI System / Use-Case Registry.} Centralized inventories documenting AI assets (common fields include: name, purpose, owner, assignee, model details, data sources, deployment status).
    
    \item \textbf{Risk Assessment Modules.} Framework-aligned evaluations that classify system risks based on data types, model providers (e.g., GPT, Claude), technical characteristics, and deployment context. These assessments are often integrated directly into use-case registries.
    
    \item \textbf{Automated ML Model Testing.} Automated tests for model fairness, robustness, accuracy, shadow-AI detection, or prompt-level safety. Some include guardrails that restrict AI behavior based on policy rules.
    
    \item \textbf{Policy Compliance Assessment.} Questionnaire- or template-based workflows aligned to individual regulations. Practitioners usually complete structured assessments for each policy (e.g., EU AI Act, NIST AI RMF), and results are stored with the documented AI asset.
\end{itemize}

Although existing platforms offer mature all-in-one governance infrastructures for enterprises, they typically keep their implementations confidential due to commercial value, limiting their contribution to research or community advancement in AI governance. Their pricing models also reflect their enterprise orientation, with many solutions requiring custom sales engagements and annual costs that place them out of reach for individual practitioners, researchers, and small organizations who typically (1) lack specialized expertise in AI policy compliance and legal requirements, and (2) operate with limited resources, making affordable automated tools an essential alternative rather than costly human evaluators (see Table ~\ref{tab:governance-tools} for pricing).

For the subset of platforms that include regulatory compliance assessments, the component most relevant to our work, we identify two dominant approaches and the gaps they introduce:

\begin{itemize}

    \item \textbf{Questionnaire-Based Compliance Assessments (LumeNova, VerifyWise, Collibra AI, Credo AI).}  
    These platforms implement compliance through policy-specific questionnaires or checklists applied to each documented AI system. After a user registers an AI use case, the platform prompts them to complete a questionnaire aligned with one specific regulation (e.g., EU AI Act, NIST AI RMF). The completed assessment is stored with the documented system and used to generate recommendations for risk classification, compliance gaps, and governance actions. Users must complete a separate form for each regulation they wish to comply with. These questionnaires translate policies into conditional survey logic, producing compliance checks that are hard-coded and inflexible. This approach leaves two gaps: (1) the lack of a flexible evaluation mechanism and (2) the absence of a single input format that can support multiple policies at once.

    \item \textbf{Code-Scanning and Automated Compliance Approaches (OneTrust, OpenLayer, Vanta).}  
    These platforms embed compliance checks into automated technical workflows, such as CI/CD pipelines or continuous monitoring systems. For example, OneTrust scans websites and applications for privacy and regulatory risks, maps them to relevant policies, and provides remediation guidance. As these automated solutions require access to system metadata or source code, they can only be used on implemented systems. This leaves a gap for flexible solutions that support compliance assessment earlier in the development process.

\end{itemize}

\subsection{AI Documentation Standards and Model Cards}

Most regulatory audits for AI systems depend on documentation as the primary source of evidence ~\cite{kierans2025catastrophicliabilitymanagingsystemic, Mokander2023, Falco2021}. However, the AI system documentation landscape is fragmented: different actors document different phases of the AI lifecycle and existing templates emphasize technical capabilities rather than legal obligations. This growing gap has increased interest in standardized documentation artifacts that encode system characteristics in structured and analyzable forms. Within this space, Model Cards and their variations have become some of the most influential approaches ~\cite{mehta2023dynamicdocumentationaisystems, 11190346, rao2025riskragdatadrivensolutionimproved}.

Model cards, lightweight Markdown documents known as ``nutrition labels for machine learning models'', offer a valuable starting point for machine learning models due to their standardized structure and ability to capture essential information, such as intended use, training data, evaluation metrics, and ethical considerations ~\cite{huggingface_model_cards_doc, Mitchell_2019}. They can be viewed as tables for documenting essential model details and are particularly convenient to use as AI practitioners can quickly create them by completing standardized templates without disclosing sensitive data, allowing for flexible information sharing based on confidentiality levels. However, the original model card framework was specifically developed for machine learning models and was intended primarily as a way to package and communicate key system information. As a result, subsequent studies have adapted the model-card paradigm to serve various domain-specific purposes.

\paragraph{The EU AI Act--Oriented Extensions of Model Cards}
Several recent extensions explicitly center on alignment with the EU AI Act. For example, Marino et al. introduced compliance Cards as a machine-readable format that encodes EU AI Act-specific metadata as an input format to a rules-based algorithm to generate compliance predictions  ~\cite{marino2024compliancecardsautomatedeu}. Similarly, Use Case Cards standardize descriptions of an AI system by combining UML diagrams with structured tables to articulate real-world scenarios and anticipate obligations under the Act’s risk-based framework ~\cite{bogucka2024usecasecards}. These artifacts implement EU AI Act requirements and are designed around that specific regulatory context.

\paragraph{Runtime Governance Cards}
Some documenting approaches focus on technical governance. Policy Cards introduce a machine-readable, deployment-layer artifact that specifies what autonomous AI agents are permitted, required, or forbidden to do at runtime. Encoded as JSON schemas, Policy Cards can be integrated with CI/CD pipelines and map operational constraints to major assurance frameworks such as the NIST AI RMF, ISO/IEC~42001, and the EU AI Act, focusing on regulating system behavior at runtime and addressing the technical enforcement layer ~\cite{https://doi.org/10.5281/zenodo.17464706}.

\paragraph{Human-Centered Enhancements to Model Cards}
A parallel line of work focuses on improving the accessibility and communicative value of documentation for a broader range of non-expert stakeholders. Interactive Model Cards and Impact Assessment Cards both enhance the usability of AI documentation by transforming static information into visually accessible, customizable, and guidance-rich formats that help diverse stakeholders interpret model behavior and societal impacts ~\cite{Crisan_2022, Bogucka_2025}. 

Existing AI governance-related model card frameworks either 1) target a single regulatory regime, 2) regulate technical behavior, or 3) focus on human interpretability. What remains missing is an accessible documentation artifact that supports multi-policy compliance assessment for a broad range of AI practitioners. PASTA fills this gap by introducing a descriptive, optimally minimal model-card format that supports fine-grained, cross-policy analysis.

\subsection{Existing Uses of Large Language Models (LLMs) in Compliance Evaluation}
AI-based compliance reasoning has long been explored, dating back to Buchanan and Headrick’s 1970s work on modeling legal inference ~\cite{buchanan1970some}. Their research examined whether AI could replicate structured legal decision processes and was furthered by the 1988 publication ``The Application of AI to Law'' ~\cite{leith1988application}, which introduced foundational symbolic and rule-based approaches to statutory interpretation. These systems established the conceptual basis for automated legal compliance revisited through modern LLMs.

Recent studies show that LLMs can achieve near--expert-level understanding of policy documents, supporting tasks such as extracting regulatory requirements, improving accuracy, and reducing manual effort in various domains ~\cite{10628489}. Ren et al. further demonstrate that the ability to process broader regulatory contexts can yield substantial gains—up to 40\% in accuracy~\cite{10628503}. Work in the HSE domain highlights the value of structured prompting: Wang et al.’s HSE-Bench uses the IRAC framework to evaluate legal reasoning and introduces Reasoning of Expert (RoE) prompts to strengthen robustness by simulating expert workflows~\cite{wang2025llmbasedhsecomplianceassessment}. Building on these advances, hybrid architectures that combine LLMs with symbolic or graph-based modules have emerged as effective compliance methods with interpretability; for example, PrivComp-KG integrates LLM retrieval with a semantic knowledge graph to align organizational privacy policies with legal frameworks, achieving a 0.9 correctness score while supporting transparent regulatory audits~\cite{10835639}.

Building on the existing body of work, PASTA extends strategies that have shown promise in LLM-based compliance evaluation, including prompting strategies, large context windows, and visual representations to support interpretable multi-policy compliance assessments.

\section{Design Goals}
Prior work has advanced automated AI governance support through various approaches. Yet our review identifies several persistent gaps. First, most questionnaire-based tools remain hard-coded to a single regulatory regime ~\cite{VerifyWise, LumenovaAI, credo2025platform, CollibraAI2024}, limiting their flexibility in today’s rapidly expanding policy landscape~\cite{oecd2023, stanford2025}. Second, code-scanning solutions are only available to implemented systems, leaving early- and mid-stage projects without support ~\cite{Openlayer2025, OneTrustWebsite, VantaWebsite, 10.1145/3685651.3686700}. Finally, commercial AI governance platforms introduce substantial financial barriers and keep their implementation details confidential (Table ~\ref{tab:governance-tools}), placing these workflows out of reach for individual practitioners and small teams.

Guided by these gaps and informed by research on AI documentation~\cite{Mitchell_2019, Bogucka_2025, marino2024compliancecardsautomatedeu} and LLM-based legal reasoning~\cite{wang2025llmbasedhsecomplianceassessment, 10835639}, we derive three design goals for PASTA:

\begin{itemize}
    \item \textbf{Design Goal 1: Scalability and Adaptability Across Multiple Diverse Policies.}  
    PASTA should evaluate heterogeneous, cross-jurisdictional policies, motivated by the rapid growth of global AI regulation ~\cite{oecd2023, stanford2025}, and the limitation of existing single-policy tools ~\cite{VerifyWise, credo2025platform}.

    \item \textbf{Design Goal 2: Interpretable and Actionable Outputs.}  
    PASTA should produce interpretable and actionable assessments that help users quickly interpret findings and translate them into actionable items. Large-scale, multi-policy evaluations generate substantial volumes of clause-level comparisons that can quickly become overwhelming, especially for practitioners without legal expertise. Prior work on model cards, interactive documentation, and explainable legal reasoning underscores the need for clear and traceable representations to support non-expert practitioner understanding ~\cite{11190346, Mitchell_2019, marino2024compliancecardsautomatedeu, Bogucka_2025}. 

    \item \textbf{Design Goal 3: Cost-Efficient Compliance.}  
    PASTA should support cost-efficient multi-policy evaluation. Existing multi-policy tools tend to be expensive, with membership prices that scale sharply as regulations accumulate (see Table ~\ref{tab:governance-tools}). Individual practitioners and small teams need an affordable, openly-available solution to navigate the complex AI policy landscape.
    \end{itemize}

By grounding PASTA in these design goals, we aim to provide an accessible, multi-policy compliance tool that complements existing governance infrastructures.

\section{PASTA: Policy Aggregator \& Scanner for Trustworthy AI}
To address our research questions, we developed PASTA and conducted assessments to evaluate its accuracy, scalability, and usability. This section details our solution.

\subsection{Motivating Usage Scenario}
To illustrate how PASTA supports real-world compliance evaluation, we present a scenario grounded in the backgrounds and challenges reported by our study participants.

Jin, a machine-learning engineer at a small research lab, is preparing an NLP prototype for deployment across North America and Europe as part of her research. Although her university’s ethics board has approved the project, she is unsure whether it complies with regional AI regulations outside her institution’s jurisdiction. Jin has moderate development experience (3--6 years) but limited familiarity with emerging AI laws. Her team lacks legal support, and previous compliance efforts relied on ad-hoc IRB forms and informal ``common-sense'' checks.

\paragraph{Step 1: Preparing System Inputs.}
Jin begins by opening the PASTA model card template in Google Sheets. The embedded instructions guide her through the required information. Within roughly 25--30 minutes, she completes the card with some information about her system.

\paragraph{Step 2: Running Multi-Policy Evaluation.}
Jin navigates to PASTA's web interface and selects AIDA, CCPA, and EU AI Act as her target regulations, reflecting her plan to release the system in North America and Europe. After clicking ``Generate,'' she waits less than a minute for PASTA to process the model card.

\paragraph{Step 3: Exploring Compliance Results.}
When the page finishes loading, Jin reviews her generated report. The overall summary indicates strong alignment across most requirements but highlights elevated risks under the EU AI Act. The interactive heatmap gives her a quick overview of compliance patterns, revealing repeated violations in Articles 6--9 that warrant attention. She then examines the policy-wise summaries for AIDA and the section-wise view to investigate specific parts of her system she is concerned about.

\paragraph{Step 4: Taking Action.}
Finally, Jin reviews the issues-and-fixes table, which provides concrete recommendations for fixing the compliance-related issues of her system.

This scenario reflects the typical workflow observed in our study: practitioners begin with limited policy awareness, complete the model card with manageable effort, and use PASTA’s summaries and visualizations to surface compliance gaps and plan targeted improvements.

\subsection{System Architecture Overview}
\begin{figure*}
    \centering
    \includegraphics[width=1\linewidth]{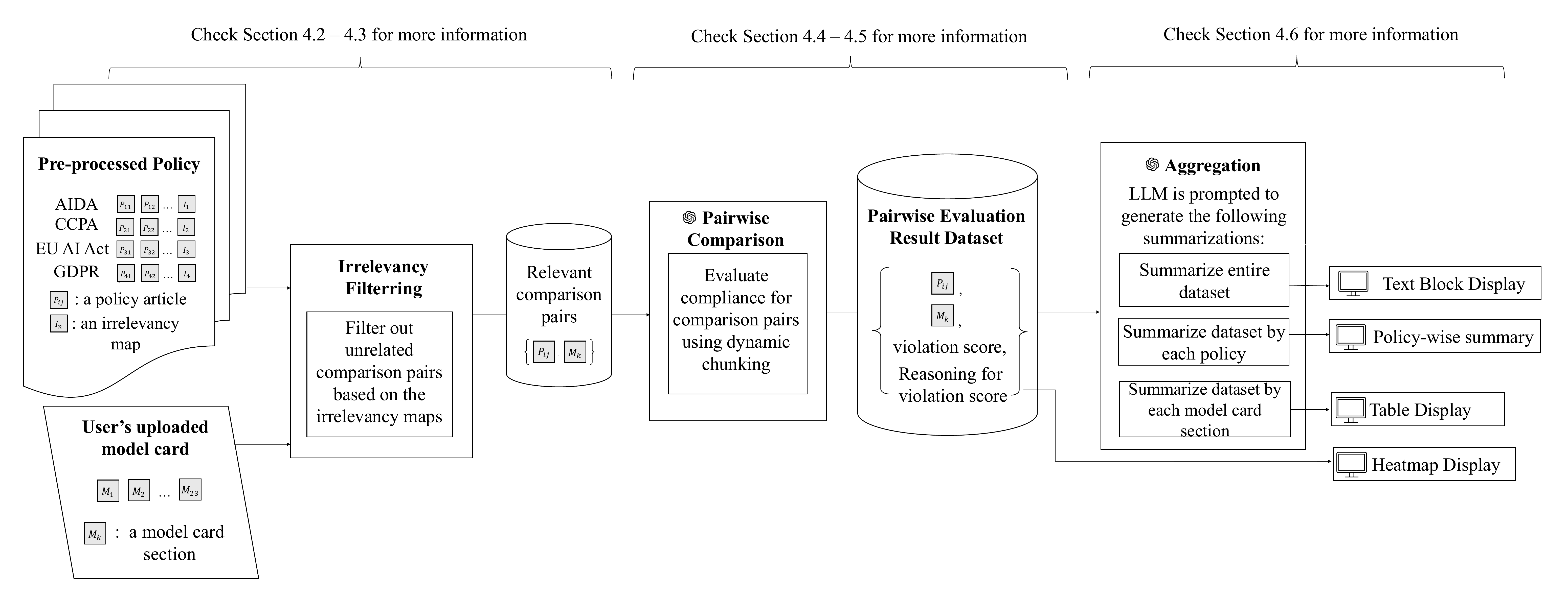}
    \caption{Overview of PASTA’s system workflow, from user-provided model card input to structured, multi-policy compliance evaluation output. The system integrates irrelevancy filtering, pairwise comparisons, violation scoring, and aggregation into interpretable outputs such as summaries, tables, and heatmaps.}
    \label{fig:pasta-workflow}
\end{figure*}

PASTA is built on a modular architecture that integrates several interconnected components to facilitate AI policy compliance evaluation (Fig. ~\ref{fig:pasta-workflow} depicts the PASTA workflow). The system features: 

\begin{itemize}
    \item A lightweight, compliance-specialized model card template that standardizes inputs.
    \item A reusable preprocessing pipeline that standardizes diverse policy documents for consistent evaluation.
    \item A granular pairwise evaluation engine powered by large language models (LLMs) that enables fine-grained comparisons between model card sections and policy articles.
    \item A multi-component web-based interface to support user-friendly interpretation of compliance evaluation outcomes.
\end{itemize}

The overall evaluation pipeline in PASTA follows a model card-to-policy comparison workflow. This process begins by taking the user-uploaded model card, structured by sections, and comparing it against preprocessed policy documents such as AIDA, CCPA, and the EU AI Act, which are segmented into articles. To optimize processing, an irrelevancy filter first eliminates unrelated pairs of model card sections and policy articles to reduce computational load. The remaining relevant pairs are then evaluated using a large language model (LLM) that performs pairwise comparisons to assess potential policy violations and generate structured reasoning. These results are compiled into a dataset, which was subsequently aggregated into human-readable summaries at multiple granularity levels. Finally, the summarized results are displayed on the user interface through interactive heatmaps, tables, and text blocks to support interpretable, actionable compliance assessments for AI practitioners.

\subsection{Designing a Model Card Template Specialized for Multi-Policy Compliance}

\begin{table}[t]
\newcolumntype{P}[1]{>{\raggedright\arraybackslash}p{#1}}

\caption{The PASTA model card structure. It captures system-level information such as intended use, compliance status, data practices, and monitoring procedures in a standardized format for policy evaluation. Check Appendix for the complete model card template with detailed descriptions for each section.}
\centering
\small
\begin{tabular}{p{3cm} p{4cm}}
\toprule
\textbf{Category} & \textbf{Section} \\
\midrule

\multirow{5}{3cm}{General Information}
& System Name \\
& Versioning Information \\
& Primary Developer/Org \\
& Contact Information \\
& System Overview \\
\midrule

\multirow{3}{3cm}{Intended Use}
& Primary Intended Uses \\
& Primary Intended Users \\
& Out-of-Scope Use Cases \\
\midrule

\multirow{2}{3cm}{Existing Compliance Information}
& Terms and Conditions \\
& Current Legal Compliance Status \\
\midrule

\multirow{4}{3cm}{System Data Information}
& Dataset Description \\
& Collection Method \\
& Bias Mitigation Measures \\
& Usage Constraints \\
\midrule

\multirow{4}{3cm}{System Performance and Evaluation}
& Summary of Performance \\
& Disaggregated Performance \\
& Testing Contexts \\
& Edge/Adversarial Testing \\
\midrule

\multirow{3}{3cm}{Ethical Considerations}
& Potential Risks and Harms \\
& Actions Taken \\
& Misuse Scenarios \\
\midrule

\multirow{2}{3cm}{Maintenance and Monitoring}
& Human Oversight \\
& Update Frequency \\
\bottomrule
\end{tabular}
\label{tab:modelcard}
\end{table}

To support our first design goal and address gaps identified in related work, we curated and assembled a specialized model-card format with three key properties:

\begin{itemize}
\item \textbf{Coverage for multi-policy evaluation:} it captures compliance-relevant information across heterogeneous policies, enabling fine-grained cross-policy comparison.
\item \textbf{Descriptive instead of code-based:} it relies on high-level, descriptive documentation rather than automated code analysis, allowing flexible assessment of non-technical dimensions such as ethics, safety, data governance, and legal compliance.
\item \textbf{Lightweight:} although comprehensive at the system level, it remains concise and feasible for non-expert AI practitioners to complete with manageable effort.
\end{itemize}
This format underpins PASTA’s input mechanism and was piloted to ensure suitability for policy-aligned compliance evaluation across a wide variety of AI systems.

In our model card design, we propose the following structuring definitions to guide the organization and processing of model card content:
\begin{itemize}
    \item \textbf{Model Card Category}: A high-level grouping that organizes related system attributes.

    \item \textbf{Model Card Section}: The smallest discrete unit within a model card. Sections provide fine-grained system information and are the primary units of analysis in subsequent stages of our framework. The choice to define sections at this level is a deliberate structuring decision that supports the modularity and scalability of our evaluation process.

    \item \textbf{Example}: The \textit{General Information} category contains sections \textit{System Name, Versioning Details, Primary Developer, Contact Information}, and \textit{System Overview}. 
    
    See Table. ~\ref{tab:modelcard} for the complete model card structure.

\end{itemize}

The structure of the PASTA model cards draws from established templates used by Hugging Face, Kaggle, and Mitchell et al., but incorporates several adaptations to better support system-level documentation and compliance analysis. These adaptations focus on aligning sections with regulatory expectations, clarifying provenance and evaluation details, and foregrounding information that is directly relevant to multi-policy compliance assessment. For example, we incorporated the following adaptations:

\begin{itemize}
    \item \textbf{General Information}: Expanded to include detailed versioning practices and compliance-aware change tracking according to the documentation requirements in the EU AI Act ~\cite{eu2024_regulation1689}.
    \item \textbf{Intended Uses}: Fully adopted from Mitchell et al.'s design of the Intended Uses section ~\cite{Mitchell_2019}.
    \item \textbf{Existing Compliance Information}: Introduces explicit sections for existing terms and conditions as well as current legal compliance status to better support compliance evaluation.
    \item \textbf{System Data Information}: Combines the Evaluation Data and Training Data sections from Mitchell et al. and the Kaggle Model Card ~\cite{Mitchell_2019, var0101_modelcards}. This unified structure is better suited for AI systems beyond machine learning models, where the distinction between training and evaluation datasets is often less meaningful. By consolidating these sections, we emphasize a system-level perspective on data provenance, collection practices, and bias mitigation, aligning more closely with the needs of compliance evaluation and regulatory transparency.

    \end{itemize}
A full description of the design decisions and section-level adaptations is provided in Appendix ~\ref{app:model-card-adaptations}.

\subsection{Policy Normalization and Preprocessing for Scalable Evaluation}
\label{sec:preprocessing}

A key challenge for multi-policy compliance is that regulations from different jurisdictions use inconsistent structures and terminology. The EU AI Act organizes requirements into chapters containing numbered articles, each subdivided into ``subparagraphs.'' Canada's AIDA instead uses numbered sections divided into ``paragraphs.'' These formatting differences make direct comparison difficult and would otherwise require custom parsing logic for each new policy.

\begin{table*}
\caption{Illustration of how raw clauses from AIDA are formatted into a structured Markdown table. Each entry specifies the article, paragraph, and content, creating a uniform representation for scalable comparison.}
\centering
\footnotesize
\begin{tabular}{l l p{10cm}}
\toprule
\textbf{Article} & \textbf{Paragraph} & \textbf{Content} \\
\midrule
1 & (1) & This Act may be cited as the Artificial Intelligence and Data Act. \\
\midrule
2 & (1) & The following definitions apply in this Act. Artificial intelligence system means a technological system that, autonomously or partly autonomously, processes data related to human activities through the use of a genetic algorithm, a neural network, machine learning or another technique in order to generate content or make decisions, recommendations or predictions. (système d'intelligence artificielle) Person includes a trust, a joint venture, a partnership, an unincorporated association and any other legal entity. (personne) Personal information has the meaning assigned by subsections 2(1) and (3) of the Consumer Privacy Protection Act. (renseignement personnel) \\
\midrule
\multirow{3}{*}{3}
& (1) & This Act does not apply with respect to a government institution as defined in section 3 of the Privacy Act. \\
& (2) & This Act does not apply with respect to a product, service or activity that is under the direction or control of (a) the Minister of National Defence; (b) the Director of the Canadian Security Intelligence Service; (c) the Chief of the Communications Security Establishment; or (d) any other person who is responsible for a federal or provincial department or agency and who is prescribed by regulation. \\
& (3) & The Governor in Council may make regulations prescribing persons for the purpose of paragraph (2)(d). \\
\bottomrule
\end{tabular}
\label{tab:Example of Processed AIDA in Standardized Table Format}
\end{table*}

\begin{figure*}
    \centering
    \includegraphics[width=1\linewidth]{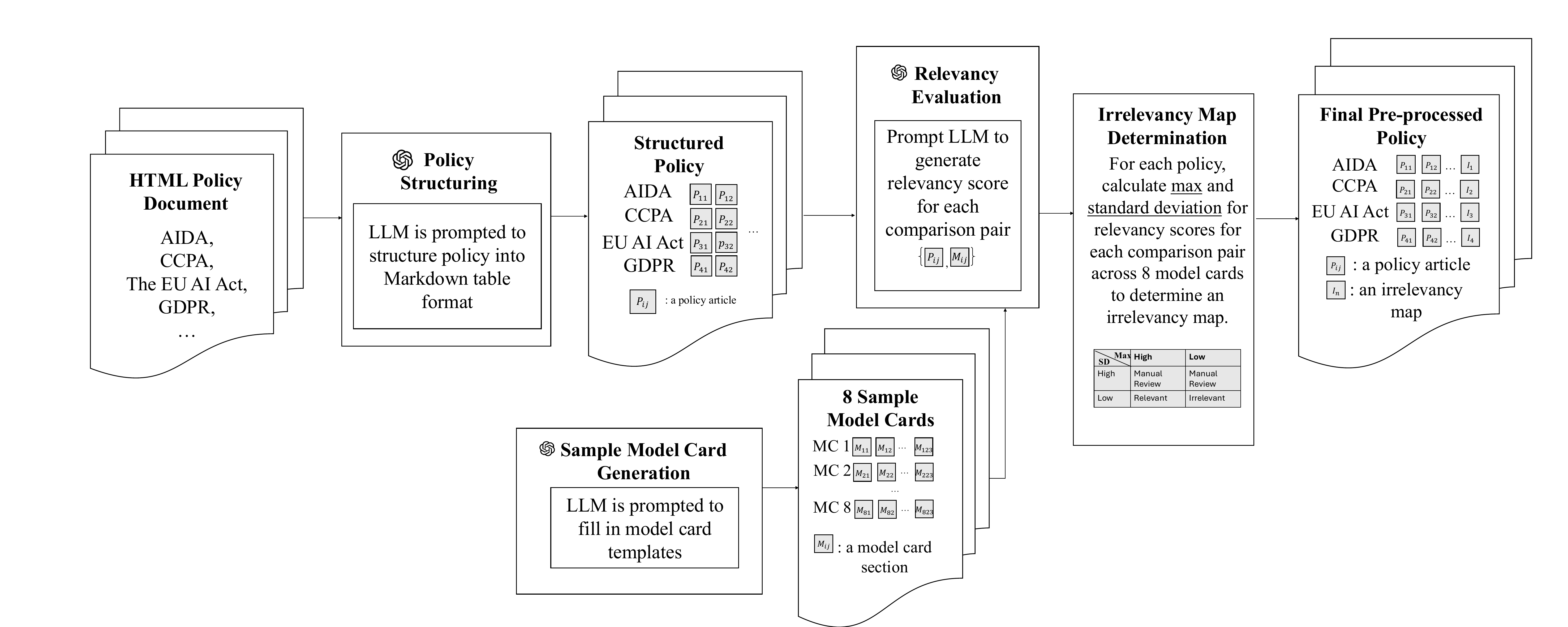}
    \caption{Illustration of how raw policy documents are segmented into paragraph-level units, standardized into a table format, and scored for relevancy against model card sections. This process generates irrelevancy maps that filter non-informative comparisons and produce a structured policy dataset for evaluation.}
    \label{fig:PASTA Policy Preprocessing Pipeline}
\end{figure*}

PASTA addresses this by normalizing all policies to a common schema. We define two structural units:
\begin{itemize}
    \item \textbf{Policy Article}: A top-level thematic block within a policy, typically numbered (e.g., Article 6, Section 12). Each article addresses a distinct regulatory requirement.
    \item \textbf{Policy Paragraph}: The smallest labeled subdivision within an article, marked by identifiers such as (1), (2), (a), or (b). Paragraphs contain the specific rules or provisions that operationalize the article's requirement.
\end{itemize}
By treating the paragraph as the atomic unit of analysis across all jurisdictions, PASTA processes any policy through a single pipeline, allowing new regulations to be added without changing the evaluation logic. We standardize policy content into a Markdown table that labels each paragraph by policy name, article number, and paragraph index, creating a uniform structure aligned with the PASTA model card and supporting scalable policy--system comparison (see Fig. ~\ref{fig:PASTA Policy Preprocessing Pipeline} and Table.~\ref{tab:Example of Processed AIDA in Standardized Table Format}). An LLM-powered preprocessing step automatically extracts and formats paragraph-level units from raw HTML, enabling consistent structuring of long, heterogeneous policy documents and ensuring reliable downstream alignment with model card content.

Not every model card section is relevant to every policy article. Comparing a section on ``Contact Information'' against an article governing risk assessment, for example, would waste computation and produce no useful signal. To avoid this, PASTA constructs an \textit{irrelevancy map} that identifies which comparisons can be safely skipped.
During preprocessing, an LLM scores each model card–policy article pair on a 0--5 scale reflecting how much that section contributes to assessing compliance with that article (see Table.~\ref{tab:CCScore} in Appendix). Pairs scoring $\leq 1$ are flagged as irrelevant and excluded from evaluation. Pairs with high score variance across sections---indicating uncertain relevance---are reserved for manual review rather than automatically discarded.
The resulting preprocessed policy package bundles the normalized policy text with its irrelevancy map. At evaluation time, this allows the pipeline to skip non-informative comparisons, reducing computational cost by 44.9\% (see Section~\ref{sec:cost-reduct}) while preserving coverage of meaningful compliance relationships.

\subsection{The Pairwise Compliance Evaluation Engine}
\label{sec:pairwise}

PASTA evaluates compliance by systematically comparing each model card section against each policy article. We call each such pairing a \textit{comparison pair}. This fine-grained structure supports Design Goal 1 (Scalability and Adaptability) by enabling the system to pinpoint exactly which aspects of an AI system's documentation comply with---or conflict with---specific regulatory requirements, regardless of policy source.

\begin{figure*}
    \centering
    \includegraphics[width=1\linewidth]{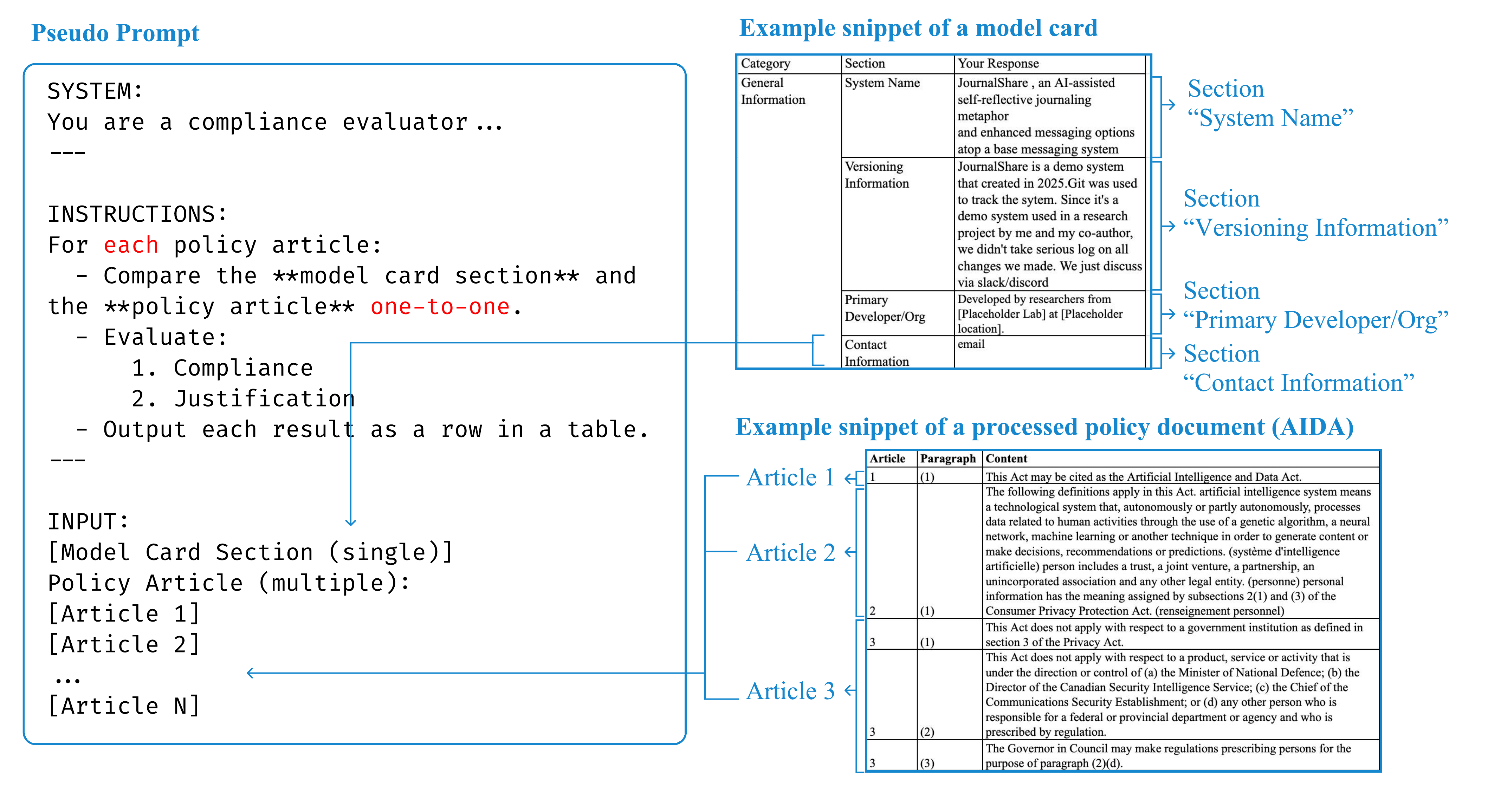}
    \caption{Illustration of pairwise evaluation: a single model-card section (e.g., ``Contact Information'') is compared against AIDA policy articles. Multiple articles are batched in one LLM request to reduce redundant calls, but the logic evaluates one section--article pair at a time, yielding one output table row per pair.}
    \label{fig:pasta-pairwise}
\end{figure*}

For each comparison pair, an LLM assesses whether the model card section satisfies, partially addresses, or violates the policy article's requirements. The full model card and policy context are cached and provided with each request. Fig. ~\ref{fig:pasta-pairwise} illustrates how comparison pairs are structured and batched within prompts. Pairs previously flagged as irrelevant (Section ~\ref{sec:preprocessing}) are skipped to reduce unnecessary computation.

We have selected Claude Sonnet 4 as the evaluation model, and the selection process is detailed in the Appendix ~\ref{app:model-selection}.

Each comparison pair receives a \textit{violation score} from 0 to 5, where 0 indicates full compliance and 5 signals a critical violation. This scale is expressive enough to distinguish minor ambiguities from serious risks while remaining intuitive for practitioners without legal training ~\cite{joshi2015likert}. See Table. ~\ref{tab:ViolationScore} in Appendix for the violation score rubric.

\subsection{Cost Optimization Strategies: Policy Chunking and Irrelevancy Filtering}
\label{sec:cost-reduct}

Evaluating a single model card against multiple policies can generate hundreds of comparison pairs. For instance, the EU AI Act alone contains over 113 articles, and Canada's AIDA spans 40. Naively issuing one LLM request per pair would incur prohibitive API overhead, while consolidating too many pairs into a single request increases hallucination risk, particularly toward the end of lengthy outputs~\cite{liu2025longcontexthallucinationdetection}. To achieve Design Goal 3 (Cost-Efficient Compliance), PASTA must balance these competing pressures: minimizing the number of requests while keeping each request short enough to maintain accuracy.

PASTA reduces costs through two complementary strategies:

\paragraph{Batching Strategy --- Grouping Semantically Similar Chunks to Minimize Redundant Requests}
Rather than grouping comparison pairs arbitrarily, PASTA uses an LLM to cluster semantically related policy articles into batches. This preserves contextual coherence---articles addressing similar regulatory themes are evaluated together---while reducing the total number of API calls. Batching semantically aligned content also improves interpretability, since related findings appear in the same evaluation pass. Through empirical testing using our selected model, we found that processing approximately 10--15 comparison pairs per request strikes an effective balance. Beyond 15 pairs, outputs become noticeably shorter and less accurate, often omitting key policy details.

\paragraph{Irrelevancy Filtering --- Identifying Unrelated Comparison Pairs to Eliminate Unnecessary Comparisons}
As described in Section~\ref{sec:preprocessing}, PASTA pre-computes an irrelevancy map that flags comparison pairs unlikely to yield meaningful compliance signals. These pairs are skipped entirely during evaluation. We also leverage prompt caching to store the full model card and preprocessed policy text, avoiding redundant input across requests.

\begin{table}
\caption{Empirical results showing how irrelevancy filtering reduces the number of comparison pairs and total runtime. 
The full evaluation compares all 23 model card sections against five policies--AIDA (40 articles), CCPA (54 articles), 
EU AI Act (113 articles), GDPR (99 articles), and the Colorado AI Act (8 articles)—resulting in 4,907 initial section–article 
pairs. Irrelevancy filtering removes pairs that consistently provide no contribution to compliance assessment, reducing 
the evaluation volume by 44.9\% without loss of coverage.}
\centering
\small
\begin{tabular}{l c}
\toprule
\textbf{Metric} & \textbf{Value} \\
\midrule
Number of comparison pairs (before filtering) & 4,907 \\
Number of comparison pairs (after filtering) & 2,204 \\
Reduction in comparison pairs & 44.9\% \\
\bottomrule
\end{tabular}
\label{tab:Evaluation Cost Reduction Achieved through Irrelevancy Filtering}
\end{table}

\subsection{Result Aggregation and Visualization: Heatmap for Overview, and Actionable Summaries for Policy-Wise and Section-Wise Understanding}

To support Design Goal 3, PASTA aggregates detailed pairwise evaluations into summaries that make large-scale compliance results easy to interpret. The evaluation process described in Section ~\ref{sec:pairwise} generates a structured dataset consisting of one evaluation result for each comparison pair. Each result includes the corresponding model card section, policy article number, violation score, and an accompanying rationale. An example of the output is provided in Appendix. ~\ref{app:Example JSON-style Compliance Evaluation Output for the System Name Section}. The results are formatted as Markdown tables and are subsequently aggregated by the LLM into interpretable compliance summaries and actionable guidance. By condensing the detailed results into targeted summaries, the system helps users efficiently understand overall compliance status, identify policy-specific gaps, and highlight system components requiring attention. The evaluation results are aggregated into the following summaries:
    \begin{itemize}
    \item \textbf{Overall Summary:} Provides a holistic assessment of the AI system's compliance across all selected policies (see interface in Fig.~\ref{fig:Example Overall Summary} in Appendix ~\ref{app:pasta-ui}).
    \item \textbf{Policy-wise Summary:} Presents a focused evaluation of the system's compliance with individual policies, offering policy-specific insights (see interface in Fig.~\ref{fig:Example Policy-wise Summary for AIDA} in Appendix ~\ref{app:pasta-ui}).
    \item \textbf{Model Card Section-wise Summary:} Identifies issues specific to each model card section and provides actionable recommendations for addressing each identified concern (see interface in Fig.~\ref{fig:Section-wise Actionable Item Table}).
    \end{itemize}

To support user interpretability, the system presents evaluation results through layered visual and textual outputs, including text block compliance summaries, an interactive heatmap of comparison pair alignment patterns, and a section-specific actionable item table.

\begin{figure*}
    \centering
    \includegraphics[width=1\linewidth]{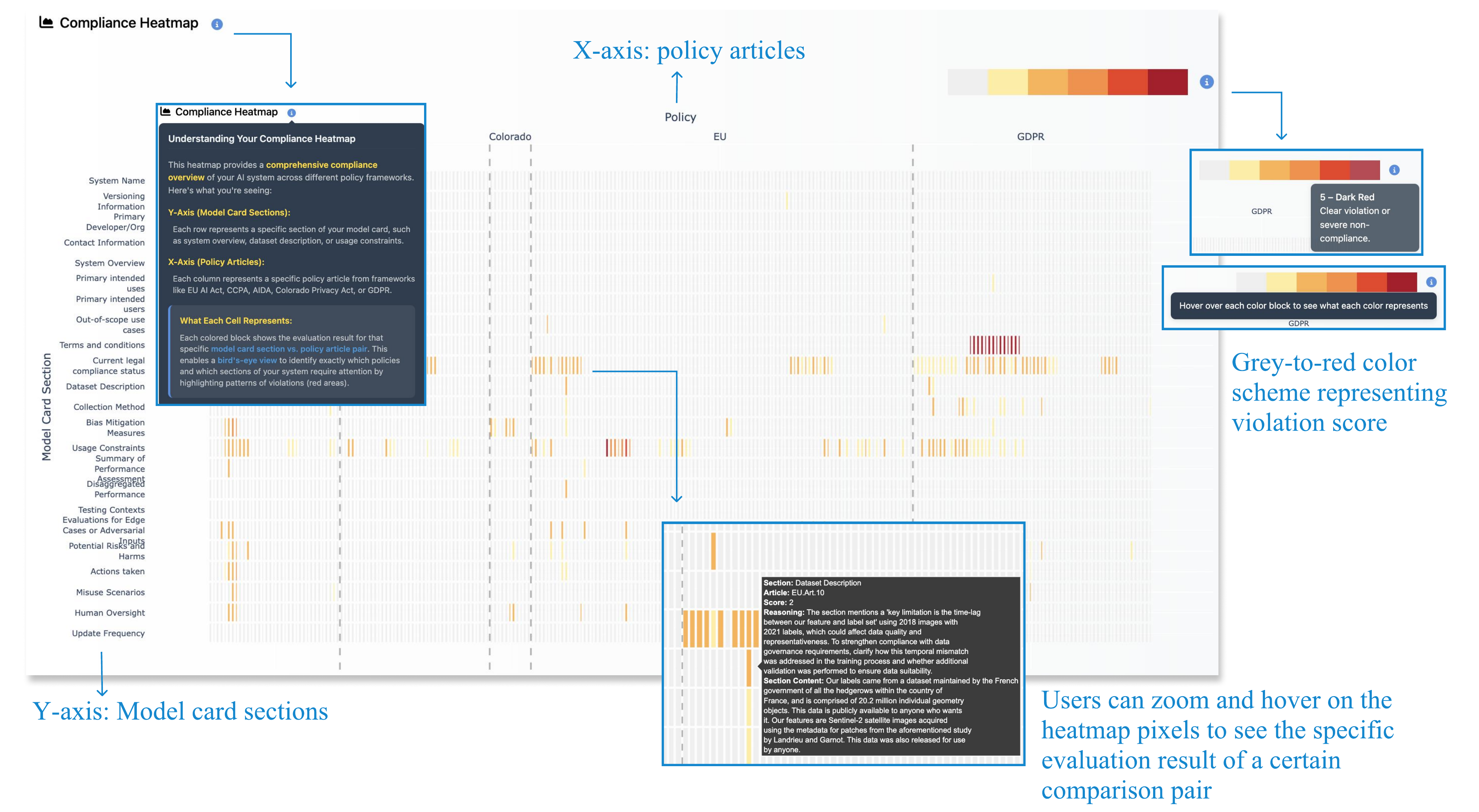}
    \caption{Heatmap interface of evaluation results across all comparison pairs. Each cell encodes violation severity, allowing users to detect patterns and compare compliance gaps across policies and system components.}
    \label{fig:Interactive Heatmap Visualization}
\end{figure*}

\begin{figure*}
    \centering
\includegraphics[width=1\linewidth]{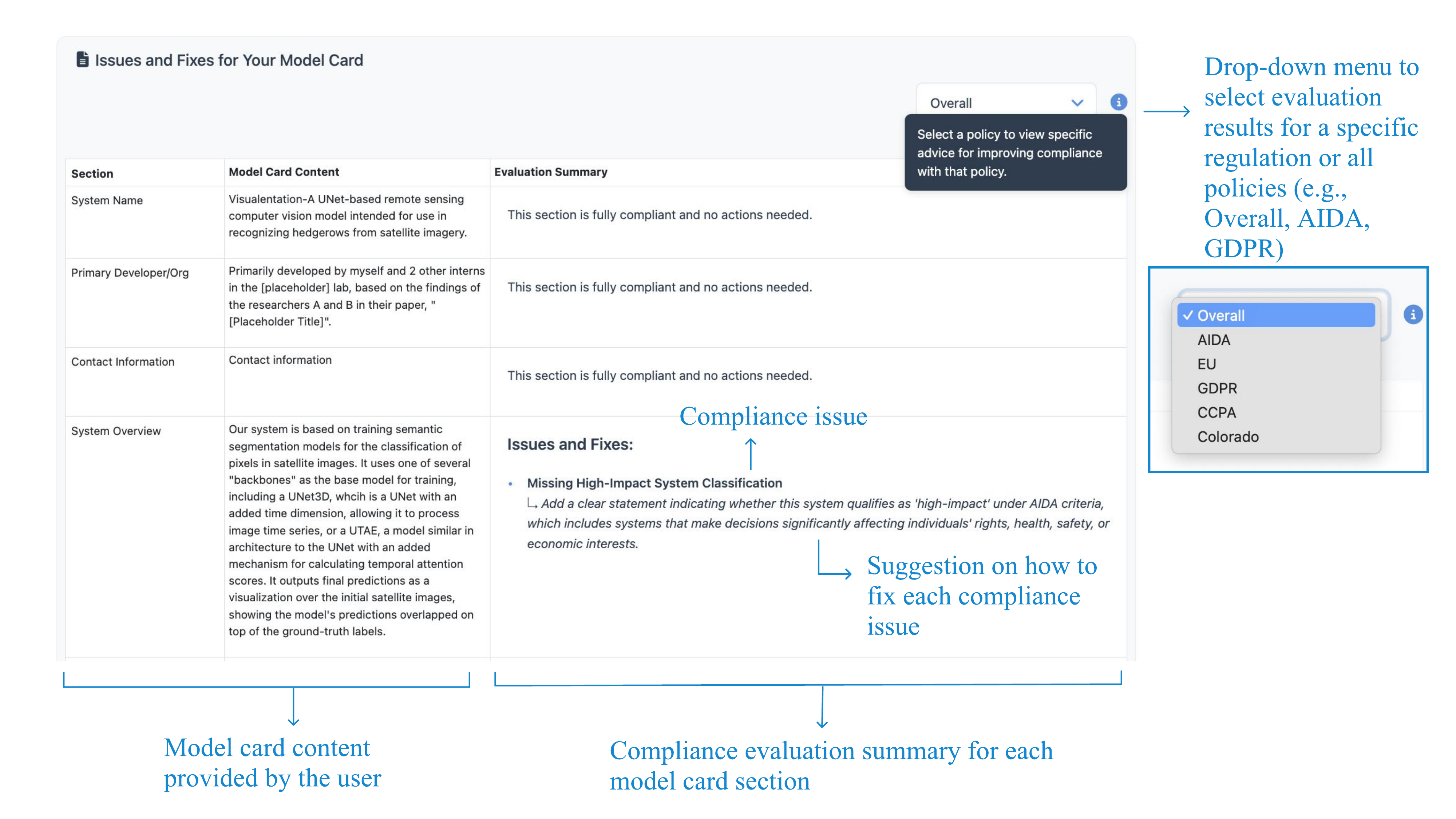}
    \caption{Section-wise Evaluation Interface linking compliance issues to specific model card sections. Each row presents the user’s documentation, detected issues, and concrete recommendations for remediation, supporting actionable system improvements. This table also supports policy-specific filtering.}
    \label{fig:Section-wise Actionable Item Table}
\end{figure*}

The interactive heatmap (see example in Fig.~\ref{fig:Interactive Heatmap Visualization}) serves as an efficient visualization tool to navigate the dense evaluation result dataset, where each data point represents the compliance evaluation result of a comparison pair. By mapping policy articles along the x-axis and model card sections along the y-axis, the heatmap enables users to quickly detect compliance patterns and identify areas of potential misalignment across multiple policies. Each block within the heatmap is color-coded based on the violation score of the corresponding comparison pair for quick visual understanding. To further support in-depth analysis, the heatmap allows users to interact with individual blocks by hovering over them to reveal detailed reasoning behind each assigned score. 

To support actionable compliance guidance, the system includes a section-wise actionable items table (example shown in Fig.~\ref{fig:Section-wise Actionable Item Table}) that directly links identified issues to concrete recommendations for improvement. This feature is designed to help AI practitioners move beyond abstract compliance scores by providing practical steps to address specific gaps. The original model card content uploaded by the user is displayed in a structured table, where each row corresponds to a particular model card section. For each section, any detected compliance issues specifically regarding that section are clearly listed alongside actionable advice to support revisions. By organizing issues and fixes in this section-by-section format, the system ensures that feedback is both traceable and easy to implement.

\section{Accuracy Assessment Against Expert Judgment}

To assess PASTA's accuracy in supporting multi-policy AI compliance evaluation, we conducted a structured expert-based study comparing PASTA's outputs against ground-truth judgments by human experts. Overall, PASTA's judgments closely aligned with expert ratings, showing strong rank correlations ($\rho = .626$ for violation; $\rho = .761$ for relevance), low error magnitudes, and agreement within one point in over 87--94\% of cases. PASTA even increased overall panel consistency when included as an additional rater.

\subsection{Methods}

We sampled representative subsets of policy articles and model-card sections and recruited three legal experts to rate each section--policy pair in terms of relevance and degree of compliance. We then applied PASTA to the same pairs and compared its scores with the experts’ consensus using agreement metrics. The study procedure was approved by the institution’s research ethics board.

\subsubsection{Expert Recruitment}

We recruited 3 AI--policy compliance experts from Upwork, a globally used freelancing platform commonly relied upon in academic research to source verified professional contractors with specialized expertise. Experts were selected based on demonstrated experience interpreting ``AI regulatory frameworks,'' and they were paid \$50 USD per hour. The 3 recruited experts reflected wide diversity in legal training. They were licensed attorneys, holding JD, LLB, and LLM degrees alongside jurisdiction-specific bar qualifications, with 6--10 years of legal practice across jurisdictions including the EU, North America, Latin America, the UK, and the Middle East. Their professional backgrounds covered privacy and data protection, AI governance and regulation, technology and cyberlaw, and commercial or corporate compliance. Each expert had prior experience assessing AI-related documentation and was familiar with major regimes such as the EU AI Act and CCPA. 

\subsubsection{Sampling of Policies and Model Cards}

We sampled 18 policy articles across the EU AI Act and CCPA to ensure thematic breadth and balanced coverage, and sampled 3 diverse model cards among the 8 used in the preprocessing procedure detailed in Section ~\ref{sec:preprocessing} . To identify high-salience model card sections, we ran PASTA across all model-card sections and selected the 6 with the highest average violation scores across both policies. This resulted in 18 model-card sections which, when paired with the 18 curated articles, yielded 324 section--policy comparison pairs as our evaluation set. A full list of selected policy articles is provided in Appendix ~\ref{app:curated-articles}.

\subsubsection{Task Procedure}

All 3 experts independently evaluated the same evaluation set. For each pair, experts provided:

\begin{itemize}
    \item a relevance score (0--5) indicating how essential and relevant the section is for assessing the policy clause;
    \item a violation score (0--5) indicating the degree of misalignment with the policy requirement. 
\end{itemize}

Experts completed the evaluation asynchronously and without time limits. Reported completion times averaged 8.7 hours ($SD = 3.5$). In a subsequent paragraph, experts supplied brief written justifications for any non-zero score, which were later used for qualitative comparison. To assess whether PASTA maintains accuracy when operating at scale, we first ran it on all three model cards against all five policies. From this full evaluation, we then selected the matching samples of the 324 section–article pairs for comparison with expert judgments.

\subsubsection{Analysis}
For each section--article pair, we constructed ground-truth labels by computing the median of the experts' ratings, and quantified agreement between experts using intraclass correlation coefficients (ICC). PASTA’s accuracy was evaluated by comparing its predictions to expert consensus using mean absolute error (MAE), Spearman rank correlation, and the proportion of cases in which PASTA’s score fell within ±1 point of the expert label. To visualize where PASTA’s predictions diverged from expert judgments, we generated confusion matrices for both violation and relevance scores and plotted scatter diagrams comparing PASTA’s outputs to consensus values. 

\subsection{Findings}

Overall, the results indicate that PASTA offers judgments that are comparably stable and human-aligned.

Before evaluating PASTA, we first assessed whether the 3 human experts formed a reliable evaluation baseline. Their intraclass correlation coefficients were .5940 for evaluating the violation score and .6393 for the relevance score, which can be interpreted as moderate-to-strong agreement based on established ICC guidelines ~\cite{article, Koo_Li_2016}. This confirms that the expert-annotated dataset provides a consistent foundation for model comparison.

\begin{table}[t]
\centering
\caption{Spearman correlations between PASTA and expert consensus, with 95\% confidence intervals.}
\label{tab:correlation}
\begin{tabular}{lcc}
\toprule
{} & \textbf{Spearman $\rho$} & \textbf{95\% CI} \\
\midrule
Violation Score & 0.6264 & [0.5553, 0.6883] \\
Relevance Score & 0.7611 & [0.7111, 0.8034] \\
\bottomrule
\end{tabular}
\end{table}

\emph{PASTA’s compliance reasoning closely aligns with those of the human experts.} Its violation and relevance scores closely mirrored expert results, reflected in Spearman correlations of $\rho = .626$ for violation scores and $\rho = .7611$ for relevance scores (see Table~\ref{tab:correlation}). Using widely cited interpretation thresholds (Cohen: .50--.69 = large; Mukaka: .60--.79 = strong), these values represent strong correlations ~\cite{Cohen1988, Mukaka2012}. Its score magnitudes remained well-calibrated: mean absolute error was low ($.284$ for violation; $.542$ for relevance score), and the system matched expert scores within one point in 94.14\% and 87.35\% of cases respectively. These patterns indicate that PASTA approximates score levels in ways that align with expert expectations.

\begin{table}[t]
\centering
\caption{Intraclass Correlation Coefficients (ICC) for human experts alone and with PASTA included as a fourth evaluator.}
\label{tab:icc}
\begin{tabular}{lcc}
\toprule
{} & \textbf{Human Only} & \textbf{Human + PASTA} \\
\midrule
Violation Score ICC & 0.5940 & 0.6249 \\
Relevance Score ICC & 0.6393 & 0.6607 \\
\bottomrule
\end{tabular}
\end{table}

\emph{PASTA increases expert agreement as its judgments are directionally aligned with human judgements.} When incorporated as a fourth evaluator in the intraclass correlation coefficients(ICC) calculations, PASTA further increased overall panel consistency, with ICC rising from .5940 to .6249 for violation score and from .6393 to .6607 for relevance score (see Table~\ref{tab:icc}). According to ICC interpretation guidelines, these values represent moderate-to-good reliability, indicating that PASTA behaves more like a stable additional rater than a source of noise ~\cite{article, Koo_Li_2016}.

\begin{figure*}
    \centering
\includegraphics[width=1\linewidth]{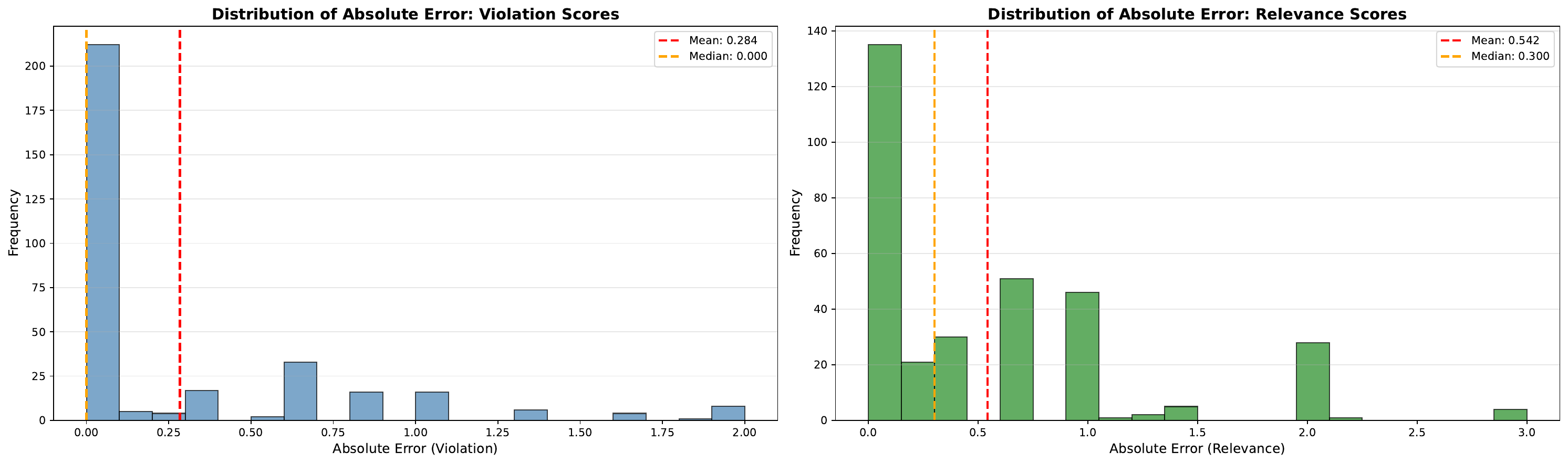}
    \caption{Distribution of absolute error between PASTA and expert consensus for (left) violation scores and (right) relevance scores. Violation errors are heavily concentrated at 0, reflecting tight alignment with experts in low-severity cases, while relevance errors span a slightly wider range, consistent with the higher variability in relevance judgments.}
    \label{fig:hist_abs_error}
\end{figure*}

\emph{PASTA demonstrated stable and directionally consistent behavior across both violation and relevance scoring tasks.} Its outputs closely tracked expert judgments in common, low-severity cases, while deviations in higher-severity violation scores followed a systematic conservative pattern—PASTA tended to slightly under-estimate severity rather than drift unpredictably. Relevance scoring spanned the full range and aligned more closely with expert distributions, reflecting that relevance judgments are easier for the model than detecting high-severity misalignment. Together, these patterns indicate that PASTA applies coherent internal criteria instead of producing noisy or erratic predictions, and that its conservative stance in high-severity cases reflects caution rather than rejection. Full confusion matrices and scatter plots illustrating these trends are provided in Appendix ~\ref{app:eval-material}.

\section{System Scalability: Runtime and Cost Evaluation}
Ensuring scalability requires assessing both PASTA’s cost of running multi-policy compliance checks and its cost of incorporating new regulatory documents. Using 3 diverse model cards among the 8 used in the preprocessing procedure, we found that evaluating 5 major policies completes in approximately 1.5–1.8 minutes at a cost of \$2.86--\$3.06, demonstrating low and stable marginal cost as policies accumulate. In Section ~\ref{sec:cost-adapt}, we further show that adapting new policies incurs a similarly lightweight cost of only 2.24 minutes and \$4.44 even for a densely structured regulation like the EU AI Act. Together, these results indicate that PASTA can scale efficiently along both dimensions without imposing substantial computational overhead. LLM models used in both Section ~\ref{sec:cost-eval} and Section ~\ref{sec:cost-adapt} are Claude Sonnet 4 and were ran on 2025-12-03.

\subsection{Scalability of Multi-Policy Evaluation}
\label{sec:cost-eval}
PASTA maintains consistently low cost per policy even when scaling to long, heterogeneous regulations. Across three full multi-policy runs on our model cards, total evaluation cost ranged from \$2.86--\$3.06, with an average of \$2.93 (SD=.10), despite substantial differences in policy length (e.g., 165{,}002 tokens for the EU AI Act vs.\ 40{,}129 tokens for Colorado). Although token usage rises with policy length, our marginal cost remains small due to irrelevancy filtering, which eliminates approximately 45\% of comparison pairs before evaluation. For example, total tokens increased by roughly 22k between Run ~1 and Run ~3, the corresponding cost increased only about \$0.20, confirming that PASTA's cost-saving mechanisms prevent runaway token inflation as the input window fluctuates.
This stability across runs suggests that PASTA's policy chunking and irrelevancy mapping strategies effectively keep marginal cost low.
\begin{table}[h]
\caption{Overall cost, runtime, and token usage for three complete multi-policy evaluation runs across AIDA, CCPA, Colorado, GDPR, and the EU AI Act. The Time column reflects the sum of evaluation and summary generation durations for each run.}
\centering
\setlength{\tabcolsep}{2pt}
\small
\begin{tabular}{lrrrrr}
\toprule
\textbf{Run} & \textbf{Total Tokens} & \textbf{Time (min)} & \textbf{Input Cost} & \textbf{Output Cost} & \textbf{Total Cost} \\
\midrule
Run 1 & 655{,}733 & 1.49 & \$1.7430 & \$1.1210 & \$2.8640 \\
Run 2 & 655{,}097 & 1.48 & \$1.7418 & \$1.1173 & \$2.8591 \\
Run 3 & 677{,}753 & 1.82 & \$1.7759 & \$1.2866 & \$3.0626 \\
\midrule
\textbf{Mean} & 662{,}861 & 1.60 & \$1.75 & \$1.17 & \$2.93 \\
\bottomrule
\end{tabular}

\end{table}

Evaluation time shows linear to sublinear scaling, completing within 1.49--1.82 minutes across runs. Because policies are evaluated in parallel, total runtime is bounded by the longest single-policy evaluation rather than by the number of policies. For example, in run 3, all 5 policies were evaluated concurrently in 40.75 seconds (0.68 minutes), followed by 68.14 seconds for summary generation. 

\begin{table}[h]
\caption{Representative per-policy breakdown for one multi-policy run (Run~1), showing input/output tokens, number of LLM requests, and total monetary cost for each regulation. Note: all policy evaluations ran concurrently; in our longest run, total evaluation time for all policies was 40.75 seconds (0.68 minutes) before aggregation.}
\centering
\setlength{\tabcolsep}{2pt}
\small
\begin{tabular}{lrrrrr}
\toprule
\textbf{Policy} & \textbf{In Tokens} & \textbf{Out Tokens} & \textbf{Requests} & \textbf{Total Cost} & \textbf{Run} \\
\midrule
AIDA        & 76{,}940  & 11{,}904 & 40 & \$0.4094 & 1 \\
CCPA        & 66{,}323  & 11{,}310 & 33 & \$0.3686 & 1 \\
Colorado    & 40{,}129  & 3{,}392  & 22 & \$0.1713 & 1 \\
EU AI Act   & 165{,}002 & 31{,}081 & 87 & \$0.9612 & 1 \\
GDPR        & 87{,}122  & 13{,}988 & 46 & \$0.4712 & 1 \\
\bottomrule
\end{tabular}
\end{table}

These findings indicate that multi-policy evaluation remains computationally efficient: cost grows slowly with policy length, execution time stays within minutes, and variance across runs is minimal. This satisfies the first dimension of scalability.

\subsection{Overhead of New Policy Adaptation}
\label{sec:cost-adapt}

PASTA's scalability also depends on the cost of integrating new regulatory documents into the system as the current policy landscape is fast-evolving. Adapting a new policy requires two automated steps described in Section~\ref{sec:preprocessing}: (1) policy structuring, where the legal text is normalized into PASTA’s standardized table format, and (2) automated relevancy evaluation, where an LLM scores each comparison pair on a 0--5 relevance scale and constructs an irrelevancy map. For the EU AI Act, policy structuring completed in 94.88 seconds and cost \$3.48, processing all API requests concurrently. Automated relevancy evaluation required only 39.47 seconds and cost \$0.96. These results show that even for one of the longest and most structurally complex AI regulations, both steps incur minimal computational cost.

PASTA’s relevancy evaluation includes an optional manual-review phase, which took 15--20 minutes on average in our earlier findings. We exclude it here to isolate the system's intrinsic scalability. Together, policy structuring and automated relevancy evaluation cost a total of \$4.44 and required 134.35 seconds (2.24 minutes), demonstrating that the pipeline can incorporate new policies rapidly and with predictable resource requirements. This fulfills the second scalability dimension by showing that PASTA can continuously expand its policy library with low marginal cost and minimal developer burden.

\begin{table}[t]
\caption{Automated cost of adapting a new policy into PASTA. Policy structuring processes the legal text into PASTA’s normalized format; automated relevancy evaluation identifies clause-level mappings to canonical requirements. Manual review is excluded because it is optional and user dependent.}
\centering
\footnotesize
\setlength{\tabcolsep}{3pt}
\begin{tabular}{>{\raggedright\arraybackslash}p{0.42\columnwidth} r r r r}
\toprule
\textbf{Processing Step} & \textbf{Time} & \textbf{Input} & \textbf{Output} & \textbf{Cost} \\
 & \textbf{(min)} & \textbf{Tokens} & \textbf{Tokens} &  \\
\midrule
Policy Structuring (EU AI Act) & 1.58 & 892{,}965 & 53{,}673 & \$3.48 \\
Automated Relevancy Evaluation & 0.66 & 281{,}148 & 7{,}994 & \$0.96 \\
\midrule
\textbf{Combined Automated Cost} & 2.24 & 1{,}174{,}113 & 61{,}667 & \$4.44 \\
\bottomrule
\end{tabular}
\end{table}

\section{Usability Evaluation}

To assess PASTA’s usability in supporting multi-policy AI compliance evaluation, we conducted a mixed-methods user study with AI practitioners who used PASTA to evaluate their AI systems spanning diverse development scopes.

\subsection{Methods}

\subsubsection{Participants}
We recruited 12 AI practitioners from universities, research labs, open-source communities, and professional teams, aligning with PASTA’s goal of supporting groups often overlooked by commercially dominant governance tools designed for well-resourced industry teams.

Our recruitment criteria included individuals over the age of 19 with at least one year of AI-related experience, such as developers, product managers, UI/UX designers, or researchers, who had worked on an AI system and were willing to use one of their systems (past or present) for evaluation in this study. These recruitment criteria accurately represent PASTA’s target user group, which are individual practitioners or small teams who build AI systems but typically lack dedicated legal expertise or compliance support.

For this study, an AI system refers to any app, model, or service that uses AI techniques---for example, machine learning models, recommendation engines, chatbots, or tools built on large language models. Importantly, the AI systems evaluated in our study spanned diverse development settings, ensuring coverage across a wide range of contexts and purposes.

We recruited 12 participants (4 women, 7 men, 1 non-binary) representing varied levels of experience and roles in AI development. Most participants had between 3--10 years of programming experience, and half had previously contributed to more than three AI systems. Their expertise spanned the development pipeline, with all serving as developers (12), alongside involvement in UI/UX design (9), project management (5), and quality assurance or testing (3). Participants reported diverse levels of familiarity with AI policies: 7 of 12 somewhat agreed they were familiar with relevant regulations, the remainder ranged from neutral (3) to low familiarity (2), indicating meaningful variation across the group. 

The systems evaluated reflected considerable diversity in both technical sophistication and application domains. Most prominently, participants brought generative models (11), chatbots (6), and deep learning models (5). Other systems included computer vision, natural language processing, recommender systems, and machine learning classifiers (3 each). These systems emerged from various development contexts, including research initiatives (10), personal projects (5), academic coursework (5), internships (4), open-source contributions (2), and industry jobs (1). The overlap in counts reflects the multi-faceted nature of modern AI systems, where practitioners often combined multiple techniques (e.g., a generative chatbot built on deep learning) within a single project.

\subsubsection{Study Design}
Our user study had two primary goals: (1) evaluating PASTA’s effectiveness in supporting AI compliance evaluation and (2) assessing its usability in terms of ease of input and clarity of outputs. These were examined through two core components—the model card input design and the compliance report interface. For the model card, we evaluated whether it balanced ease of completion with sufficient detail for compliance assessment. For the interface, we evaluated whether it was user-friendly and supported effective interpretation. During the user study, participants engaged with PASTA in a realistic workflow: preparing inputs, generating reports, and navigating results. We collected data through surveys, targeted questions, and think-aloud protocols and applied thematic analysis to capture findings and reasoning for our primary goals.

The study proceeded in five stages:

\begin{enumerate}
\item \emph{Demographic Survey (5 minutes)}: Participants reported their role, experience, and familiarity with AI policy, along with the context of the AI system evaluated using PASTA.
\item \emph{Task 1: Preparing Inputs ($\leq$ 30 minutes)}: Participants completed a model card for their AI system using a provided template and answered questions on its comprehensiveness and the effort required.
\item \emph{Interim Interview (10 minutes)}: Participants answered open-ended questions about their past compliance experiences and their perceptions of the model card approach.
\item \emph{Task 2: Report Exploration (15--20 minutes)}: Participants used the PASTA web interface to review a compliance report generated from their model card, during which they identified compliance issues and potential solutions. Think-aloud protocols and open-ended questions probed their perceptions of PASTA’s usability and effectiveness in supporting compliance at scale.
\item \emph{Post-Task Survey (10 minutes)}: Participants completed the System Usability Scale (SUS), the NASA Task Load Index (NASA-TLX), and Likert-scale items assessing both the model card process and the PASTA interface ~\cite{articleSUS, HART1988139}.
\end{enumerate}

 Across tasks, both quantitative (e.g., completion times, Likert ratings, and NASA TLX scores) and qualitative (e.g., interview reflections and open-ended survey responses) data were collected, allowing us to assess not only whether participants could complete model cards and interpret reports with a reasonable amount of effort but also how their understanding and attitudes evolved after interacting with the system. See our supplementary materials for the survey questions. We conducted a pilot study with four participants to refine the tasks and survey instruments. The results reported in this paper are from the subsequent 12 participants in the main study.

\subsubsection{Analysis}
Our study employed a mixed-methods design with a qualitative emphasis and quantitative support. The primary analysis was qualitative: we conducted a thematic analysis of think-aloud transcripts and open-ended responses to understand how participants engaged with the model card input and the compliance report. Following Braun and Clarke’s six-step process, we familiarized ourselves with the transcripts, generated initial codes, searched for candidate themes, reviewed and refined them, and produced final thematic descriptions ~\cite{article}. The resulting themes surfaced participants’ perceptions of the model card input design, the compliance report interface, and PASTA’s overall performance in supporting AI compliance evaluation.

Quantitative analyses complemented these findings by providing structured measures of usability and effort. We calculated descriptive statistics for the System Usability Scale (SUS), NASA TLX cognitive effort scores, and Likert-scale items, and compared SUS scores against the industry benchmark of 68, representing average usability. These numerical results helped triangulate the qualitative themes, ensuring that statistical trends informed our findings.
\subsection{Findings}
\begin{figure*}[t]
\centering

\begin{subfigure}[t]{0.24\textwidth}
    \includegraphics[width=\linewidth]{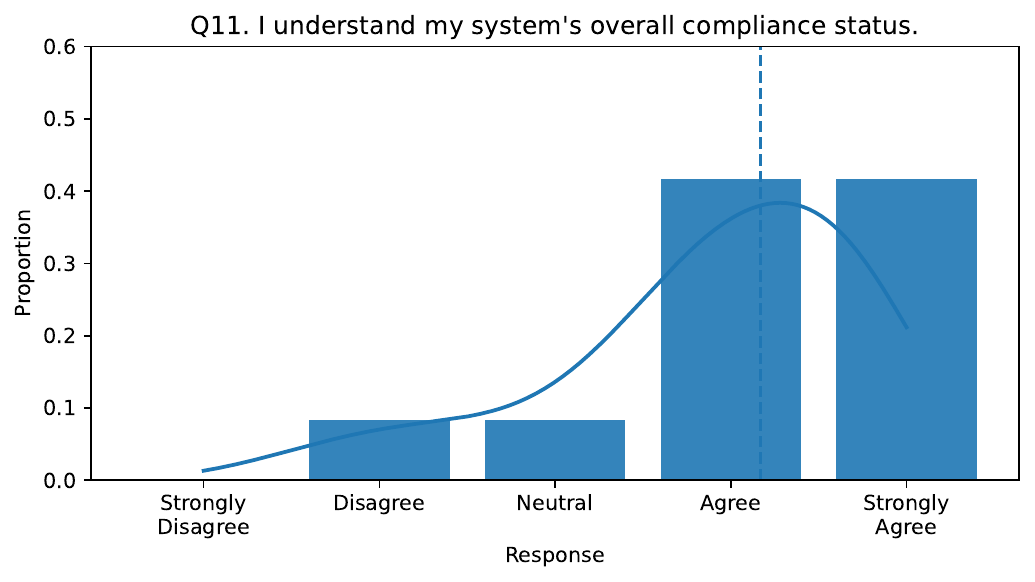}
\end{subfigure}
\hfill
\begin{subfigure}[t]{0.24\textwidth}
    \includegraphics[width=\linewidth]{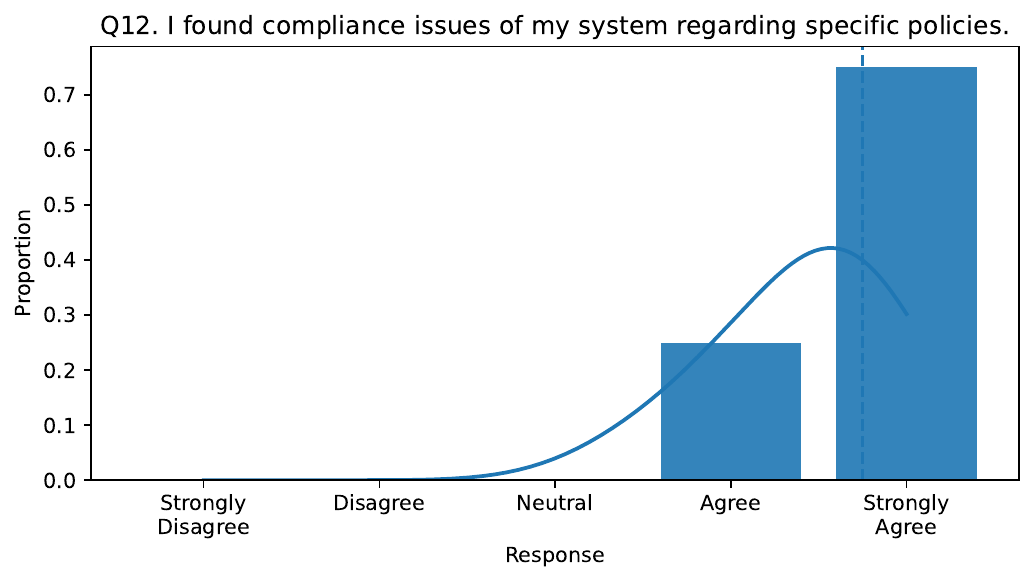}
\end{subfigure}
\hfill
\begin{subfigure}[t]{0.24\textwidth}
    \includegraphics[width=\linewidth]{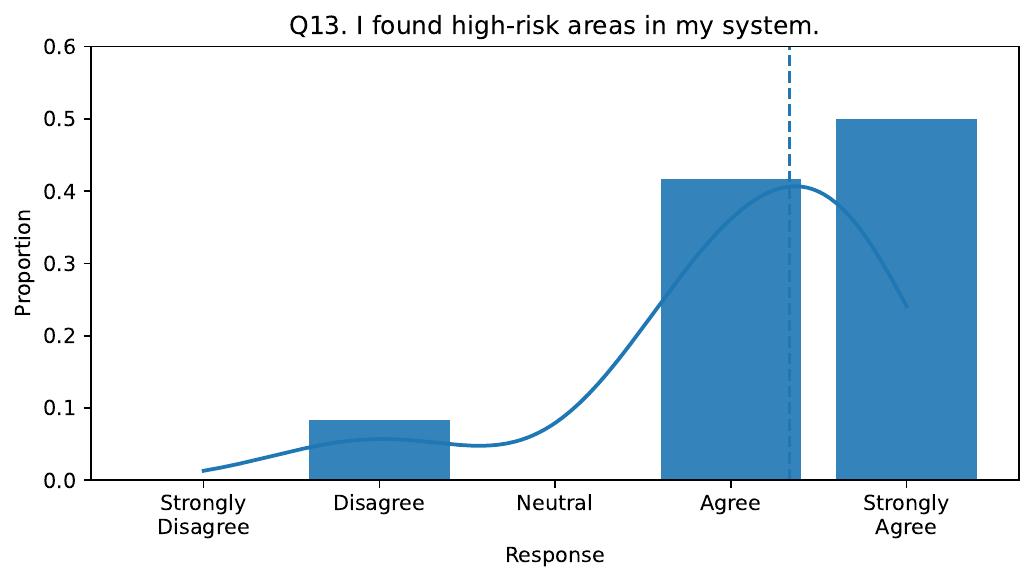}
\end{subfigure}
\hfill
\begin{subfigure}[t]{0.24\textwidth}
    \includegraphics[width=\linewidth]{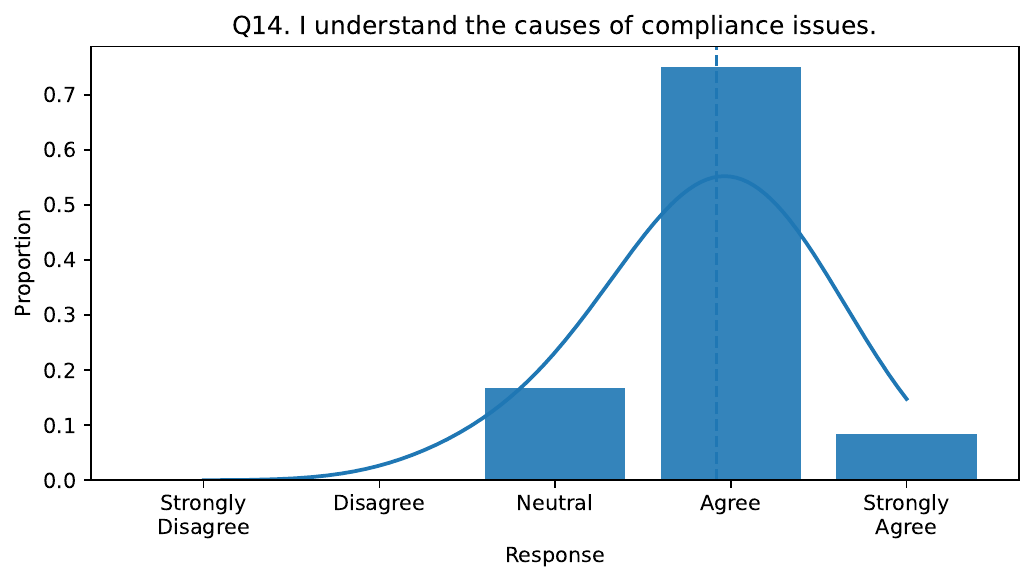}
\end{subfigure}

\vspace{0.6em}

\begin{subfigure}[t]{0.24\textwidth}
    \includegraphics[width=\linewidth]{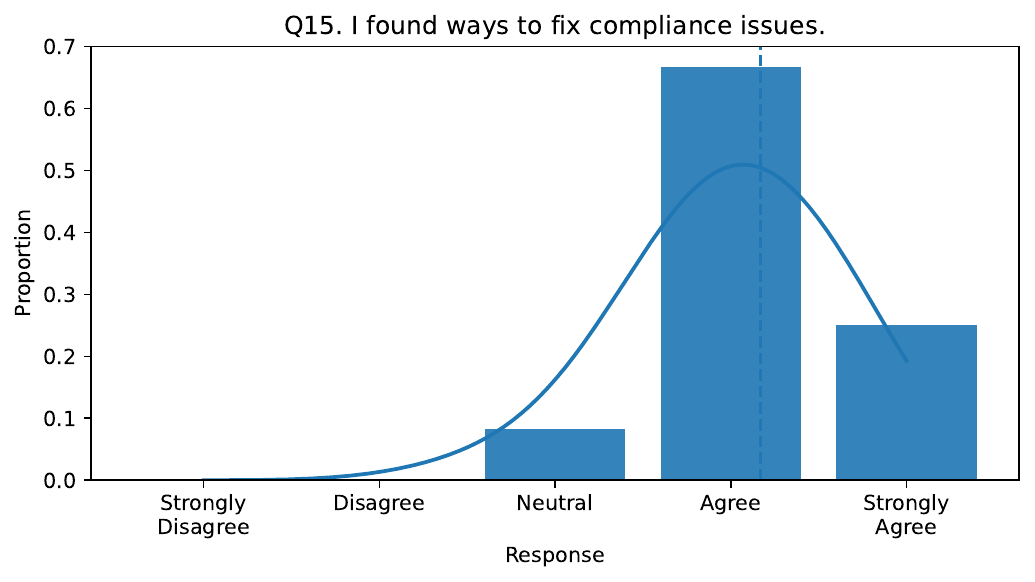}
\end{subfigure}
\hfill
\begin{subfigure}[t]{0.24\textwidth}
    \includegraphics[width=\linewidth]{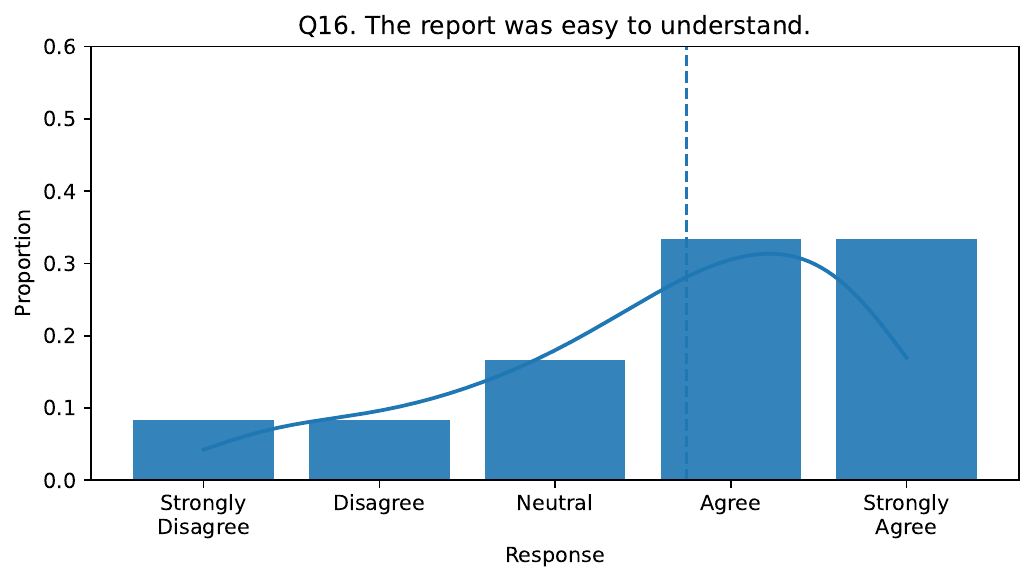}
\end{subfigure}
\hfill
\begin{subfigure}[t]{0.24\textwidth}
    \includegraphics[width=\linewidth]{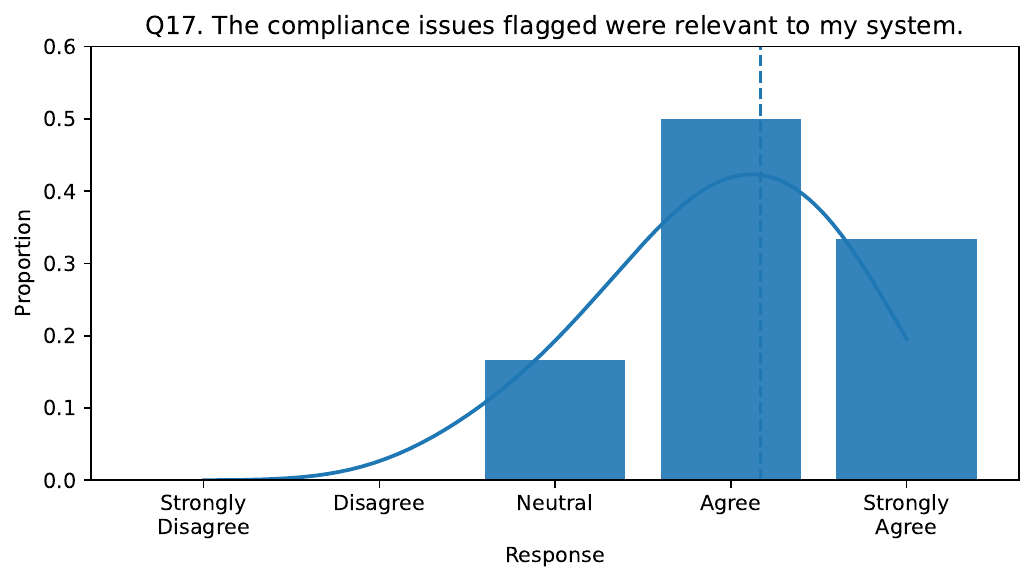}
\end{subfigure}

\caption{Response distributions for seven Likert-scale questions assessing participants’ understanding, detection, and resolution of compliance issues. Bars represent the proportion of responses and vertical lines indicate the mean (solid) with 95\% confidence intervals (dashed). Participants generally reported strong agreement in understanding their system’s compliance status (Q11), identifying specific policy issues (Q12), locating high-risk areas (Q13), and finding ways to resolve issues (Q15). Moderate agreement was observed for understanding issue causes (Q14), report clarity (Q16), and issue relevance (Q17).}
\label{fig:figure10}
\end{figure*}

Overall, PASTA proved usable (SUS $M = 73.5$, $SD = 15.2$) and effective in supporting multi-policy AI compliance evaluation. Participants described the model card input as effortful yet manageable, often framing it as a useful reflection on their systems. The output reports were seen as actionable and interpretable, and participants completed report reviews quickly, averaging 6.5 minutes ($SD = 5.72$) when surfacing policy-specific issues and linking them to system components, though they wanted stronger guidance on remediation. The tool enabled participants to identify compliance risks quickly and with confidence. Fig. ~\ref{fig:figure10} summarizes participants’ responses to the usability questions, highlighting strong agreement on the tool’s clarity, actionability, and overall usefulness. We present these findings across three themes: the effort and value of input, the clarity and actionability of output, and the tool’s impact on practitioners’ overall understanding of compliance. 

\subsubsection{The model card-based PASTA input enabled comprehensive coverage and reflective documentation with manageable effort}

Completing the model card functioned less like filling out a bureaucratic form and more like engaging in a structured reflection on the system. Although NASA-TLX ratings indicated consistently high mental demand (70--95; $M = 76.7, SD = 25.4$) and notable effort, participants framed this cognitive load as purposeful rather than burdensome. Instead of slowing them down, the card helped consolidate information they typically held in scattered notes, old IRB documents, or ad--hoc descriptions. As one participant put it, ``It felt like a check-up on my system rather than random paperwork’’ (P3). Participants consistently described the model card as a productive prompt for articulating decisions, assumptions, and risks that had never been written down in one place. This suggests that the model card acted as a reflective scaffold.

The task was also relatively time-efficient compared to traditional compliance workflows and existing AI compliance tools, illustrating that the model card served as a structured shortcut rather than an added burden. Most participants completed the card in 20--30 minutes ($M = 28.4$, $SD = 6.7$; range 19--39.5), and several contrasted this with the multi-hour process of assembling documentation for legal consultations or governance reviews. For practitioners without access to enterprise governance tools or formal legal support, this consolidation of work was especially valuable. This comparatively low-effort input also stands out when benchmarked against existing compliance-oriented tools. Platforms offering ``regulatory compliance checking’’ (such as LumeNova, VerifyWise, Collibra AI, and Credo AI; Section ~2.1) rely on questionnaire--style inputs that request information similar in scope—system overviews, oversight mechanisms, misuse scenarios—but require this work to be repeated for each regulatory regime, causing effort to multiply as policies increase. Other tool categories offer no meaningful input-time baseline: enterprise governance platforms require multi-week technical integration; risk-prediction and policy-interpretation tools accept short free-text descriptions and do not perform compliance checks; and behavior-verification tools require source code or dataset access. Even model-card variants aligned with the EU AI Act, runtime governance, or human-centered communication do not report input--time measurements and target narrower goals. Notably, Tauqeer et al.’s GDPR knowledge-graph checker reports 3.55--11.48 minutes merely to encode a \emph{single} contract instance for one policy. In contrast, PASTA’s $\sim$30-minute model-card input produces comprehensive, system-level documentation that supports \textit{multi-policy} assessment (EU AI Act, AIDA, CCPA) in a single pass. Together, these comparisons indicate that PASTA offers a favorable effort-to-coverage ratio: it concentrates focused cognitive work into a reusable documentation artifact while avoiding the compounded burdens seen in existing solutions.

Importantly, the model card surfaced gaps in understanding without becoming a bottleneck. Participants rated the instructions as clear ($M = 4.17, SD = 0.94$) and the captured information as comprehensive ($M = 4.75, SD = 0.45$), and the open-text design allowed both concise bullet points and detailed narratives. Many described moments where writing forced them to confront previously unarticulated questions: ``It made me go back through my code and documentation to check what was missing’’ (P1). Others noted that prompts about risks and harms helped them express latent concerns they had only implicitly considered (P3). Taken together, these patterns indicate that PASTA’s input requires active cognitive engagement but ultimately supports systematic thinking and reflective documentation—offering a manageable and productive pathway for preparing a multi-policy evaluation rather than introducing additional overhead.

\subsubsection{PASTA reorganized large multi-policy evaluations into decision-ready, component-level guidance}

On the output side, PASTA largely achieved its goal of making large--scale, cross--jurisdiction evaluations interpretable and actionable (Design Goals 1 and 2). Participants reported strong confidence in their ability to surface problems across different policies ($M = 4.73, SD = 0.45$) and to see where risks were concentrated. Rather than treating policies as separate checklists, they used PASTA to build a comparative picture of deployment options. One participant explained, ``I could check what problems might occur if I released a product in the U.S. That was very clear'' (P2), while another described using the reports to reason about launch order: ``It made it obvious that the EU is stricter; if resources are tight, we might start in California or Canada first'' (P10). The tool thus helped translate a heterogeneous policy landscape into a configuration of practical choices.

PASTA also helped participants connect abstract policy clauses to specific parts of their systems, though support for remediation remained only partial. Survey results indicated high confidence in locating compliance issues within system components ($M = 4.27$) and understanding why those issues arose ($M = 4.18, SD = 0.57$). Participants emphasized that the clause-linked issue descriptions ``finally told me which piece of the pipeline each requirement was about'' (P4). However, ratings for finding clear ways to fix issues were lower, though still moderate ($M = 3.91, SD = 1.16$). The issues-and-fixes table was frequently described as the most actionable element -- ``The issues-and-fixes table gave me concrete next steps\ldots that’s what I really want to know: what do I need to fix?'' (P7) -- yet several participants asked for more structure in the recommendations, such as tagging fixes by effort level or affected subsystem. This pattern suggests that PASTA effectively supports sensemaking and diagnosis, while remediation planning remains an area for refinement.

The heatmap visualization played a distinctive role in helping participants triage attention across the large result space created by scalable, multi-policy evaluation. Rather than reading every table in detail, participants used the heatmap to identify clusters of concern and prioritize limited resources: ``It visually prioritized issues across rows and columns\ldots it showed me what to fix first if time and resources are limited'' (P10). Others remarked that it surfaced patterns they would not have noticed from text alone, such as specific articles that repeatedly caused problems across jurisdictions: ``Seeing Articles 6--9 light up made it clear those are consistently problematic'' (P9). At the same time, comments requesting clearer legends, zoomable views, or shorter explanations indicate that while the heatmap succeeds as a high-level triage tool, its visual design could better support quick interpretation when the number of policies and sections grows. Overall, these findings show that PASTA’s output design enables practitioners to work with large volumes of evaluation results without being overwhelmed, even as it highlights remaining opportunities to strengthen remediation guidance.

\subsubsection{PASTA shifted practitioners’ conception of compliance from vague ethics to concrete, multi--policy obligations}

Beyond immediate usability, interacting with PASTA changed how participants conceptualized AI compliance for their own systems. Prior to the study, most framed compliance in terms of ``common sense'' ethics, IRB paperwork, or vendor requirements (e.g., ``IRB is my main encounter,'' ``it felt like the Wild West''), and only one had ever produced formal legal documentation. Even among the 7 participants who self-reported as familiar with AI policies, 5 still overlooked major obligations that PASTA later surfaced. This pattern suggests that perceived familiarity did not translate into a detailed grasp of concrete duties, especially across multiple jurisdictions.

Using PASTA made these obligations more visible and actionable. After working through the reports, half of our participants identified obligations they had not previously considered, such as dataset registration, risk-mitigation duties, and clarifying misuse and oversight. Several expressed surprise at the breadth and specificity of existing regulation: ``I didn’t realize there were already so many policies, and that they care about things like logging and appeals in this much detail.'' (P6) Five reported gaining ``a clearer understanding of what compliance means in practice,'' describing previously abstract requirements as ``visible and concrete.'' By the end of the study, more than half of participants reported feeling confident addressing most flagged issues (approximately 70--80\%), even if they still anticipated needing legal review for high-stakes deployments.

Taken together, these contrasts indicate that the main barrier for our participants was limited awareness and articulation of obligations, not lack of concern or intent. PASTA did not replace legal expertise, but it gave practitioners a working vocabulary of their responsibilities across multiple policies, making it easier to recognize where they were already aligned and where further work or expert consultation was needed. For unsupported individual developers in particular, this shift from vague ethical intuition to concrete, multi-policy obligations is itself a central usability outcome.

\section{Discussion}

In this section, we position our work in a broader perspective and discuss the wider applications and implications of PASTA across three domains: (1) strengthening documentation and harmonization practices, (2) supporting real-world compliance workflows and navigating regulatory complexity, and (3) enabling safe, collaborative, and educational use of AI-assisted compliance tools.

\subsection{Documentation and Harmonization at Scale}

PASTA's compliance-oriented model card can serve as a valuable artifact for AI governance and system documentation. In our usability study, several participants described it as a checklist that helped them surface compliance considerations they had not previously thought about. This suggests that the PASTA model card could itself function as a compliance tool that guides users through key responsibilities in developing RAI. While existing AI documentation tends to emphasize either communicative simplicity or technical depth, PASTA model card strikes a balanced design: it is compliance-focused, easy to complete for the information it captures, easy to understand as a checklist, and flexible enough for practitioners to adjust the level of detail based on their needs. This balance expands its target users from only machine learning experts or AI developers to a broader range of AI practitioners, to document and reflect on their system’s RAI posture. We believe the PASTA model card has strong potential to become a reusable compliance artifact and a practical complement to formal governance processes.

PASTA also provides a method for harmonizing legal texts through its policy preprocessing pipeline and a unified paragraph-level schema suitable for automated analysis. Differences in structure, terminology, and granularity across jurisdictions have long complicated cross-policy reasoning, not only for AI regulations but for legal documents more broadly. Our results show that PASTA’s normalization pipeline effectively reduces this fragmentation by converting diverse regulations into a consistent format, which enabled clause-level alignment. Beyond standardizing legal documents, this unifying method also supports broader regulatory analytics, such as detecting overlaps or identifying gaps.

PASTA’s pairwise comparison engine provides cost-saving strategies that can inform future large-context comparison workflows. Its scalability results show that LLM-driven text comparison and reasoning can remain efficient while maintaining accuracy. In future systems that need to compare large text corpora, irrelevancy filtering can be applied to remove non-informative segments, as our results show this substantially reduces evaluation volume without losing coverage. Likewise, policy chunking can be used to divide long documents into coherent segments to lower computational overhead while still preserving enough context for accuracy. Together, these techniques provide scalable design principles applicable to many domains that require large-context comparison, such as cross-statute harmonization in law, plagiarism detection for writing, and the matching clinical guidelines across medical standards.

\subsection{Supporting Real-World Workflows and Navigating Regulatory Complexity}

Our usability study unexpectedly revealed that PASTA integrates naturally into early--mid stages of AI system development workflow. Several participants in our usability study described it as a practical mechanism for identifying compliance obligations and missing safeguards while still exploring feasibility or refining a partially built system. Unlike prompt-based risk-prediction tools that operate at a high level, and unlike code-dependent audit tools that assume a finished product, PASTA introduces an AI compliance framework that occupies a middle ground—specific enough to surface actionable issues yet flexible enough to support iterative development or early-stage experimentation. 

PASTA’s multi-dimensional design can support small, multi-role teams by providing views that align with needs of different roles. The model card functions as an authoring artifact for audits, while the high-level summaries and heatmap offer an accessible entry point for developers and designers without legal expertise. The heatmap and issue--fix tables also help product managers and stakeholders identify priorities and translate findings into concrete tasks. Participants further noted that PASTA’s structured outputs can be integrated into ticketing or audit systems, enabling distributed teams to coordinate around standardized evidence to better support larger team settings.

The system also serves as a software-architectural model for compliance tools seeking to adapt to future regulatory change, which is fairly meaningful in today’s fast-evolving policy landscape. As both the preprocessing pipeline and evaluation logic treat policies as modular inputs, newly introduced or amended regulations can be incorporated without redesigning the framework, corresponding to our first design goal regarding scalability. This architecture naturally supports governance tasks such as policy versioning and longitudinal compliance tracking. The ability to ``plug in'' future or customized policies is essential for AI systems operating across multiple jurisdictions. While commercial solutions often adopt similar modular designs, we openly share our implementation details, making this policy adaptability available for community-driven improvements. We note that our technical evaluation assesses whether PASTA’s outputs are directionally aligned with expert judgment, rather than whether they meet a definitive standard of legal sufficiency. In practice, what constitutes ``good enough'' compliance reasoning depends on deployment context, organizational risk tolerance, and regulatory stakes, reinforcing that PASTA is designed to support decision-making rather than replace formal legal review.

\subsection{Safe and Educational Uses of PASTA}

Introducing AI into compliance work also presents risks that must be acknowledged. LLM-generated feedback can be misused to surface regulatory loopholes or encourage rule gaming—for example, one participant noted that if PASTA flagged something as problematic, they might simply remove it from their documentation. Automated compliance checks may also lead to over-reliance on model interpretations, reflecting concerns about hallucination in LLM-assisted workflows. PASTA mitigates some of these risks by grounding its outputs in clause-level justifications and adopting conservative scoring patterns, while it must remain a decision-support tool rather than a complete replacement for legal or policy review. Responsible deployment therefore requires human oversight, audits, or organizational guardrails.

PASTA also surfaced unanticipated value as an educational tool. One participant, a university instructor, noted that the system could serve as a clear and approachable introduction to AI policy and RAI in academic or professional training contexts. They emphasized that the tool’s interactive nature creates a realistic and engaging way for learners to explore policies and AI governance. Students can input their own AI systems, which increases motivation as feedback directly concerns their work. The hands-on process can also help them understand policies in depth on a practical level. By adjusting the model card content, students can experiment with policy thresholds, seeing what counts as compliant, what constitutes a violation, and whether missing information is flagged or ignored. They can also examine how different policies respond to the same system, learning the specifics of policies and how they differ. Additionally, by varying the level of detail in their descriptions, learners gain a clearer understanding of the depth and clarity expected in AI compliance audits, reinforcing broader principles emphasized in AI governance. This makes PASTA potentially useful for teaching computer science students, AI developers, and even corporate employees, as organizations could upload their own internal policies for training purposes. These emergent uses highlight opportunities to extend PASTA into instructional settings, expanding its role beyond compliance evaluation alone.

\section{Future Work}

Our study evaluated the accuracy and usability of PASTA with a relatively small number of participants and experts. Future iterations should incorporate larger-scale expert validation to more comprehensively assess its reliability. Engaging legal scholars and regulatory professionals would allow us to benchmark PASTA’s outputs against expert-annotated ground truth, providing a clearer picture of its precision and recall across multiple policies. Such validation could also surface edge cases where policy language resists straightforward alignment, informing refinements to our preprocessing pipeline and evaluation prompts.

PASTA may be extended from policy-specific evaluation toward true cross-policy reasoning. PASTA currently addresses regulatory ambiguity in multi-jurisdictional environments by keeping evaluations strictly policy-specific. Each output view allows users to inspect results for an individual policy, and each LLM request in the comparison pipeline operates on one policy at a time. This separation prevents premature blending or collapsing of obligations that may diverge or conflict across jurisdictions—a key concern in comparative policy research. However, future extensions could incorporate cross-policy conflict detection, obligation clustering, and prioritization support. These capabilities are increasingly important as organizations deploy systems across regions with rapidly evolving and sometimes inconsistent regulations and may also support policymakers seeking clearer visibility into regulatory overlaps and gaps.

Beyond accuracy, participants’ reflections revealed several new directions for extending PASTA into packaged solutions to AI compliance. Some envisioned PASTA as an educational tool for universities or professional training programs, where interactive model cards could help students and practitioners learn compliance obligations through hands-on, customized exploration rather than passive lectures. Others proposed integrating PASTA into corporate workflows; for instance, linking compliance issues to ticketing systems, expanding high-level violations into sub-tasks, and assigning them to appropriate teams. Some participants expressed interest in more detailed guidance, suggesting a chatbot interface providing step-by-step remediation advice. Together, these ideas paint a future where PASTA evolves from a static evaluation tool into a dynamic ecosystem supporting education, organizational governance, and real-time compliance assistance.

\section{Conclusion}

This work introduces PASTA (Policy Aggregator \& Scanner for Trustworthy AI), a scalable and cost-efficient compliance evaluation tool that addresses critical gaps in the current AI governance landscape. By leveraging large language models and implementing strategic design choices—including a specialized model card input format, policy structuring framework, and efficiency optimization techniques—PASTA enables AI practitioners to evaluate their systems against multiple global policies simultaneously. The findings reveal that PASTA's lightweight input design, interpretable visualizations, and actionable feedback mechanisms can bridge the gap between complex regulatory requirements and practical implementation needs.

As AI technologies continue to proliferate across industries and jurisdictions, tools like PASTA become essential infrastructure for RAI development. By democratizing access to compliance evaluation and reducing the traditionally high costs associated with legal expertise and manual assessment, PASTA empowers a broader range of organizations—from startups to research institutions—to proactively address regulatory requirements. Looking forward, the modular architecture of PASTA provides a foundation for expanding policy coverage, incorporating emerging regulations, and adapting to evolving compliance landscapes. This work represents a step toward creating an ecosystem where compliance evaluation is not a barrier to innovation but rather an integral, accessible component of the AI development lifecycle, ultimately fostering greater trust, accountability, and responsible deployment of AI systems globally.

\begin{acks}
This research was supported by the Korea Institute of Science and Technology (KIST) institutional program (26E0062). We sincerely appreciate their support, which made this work possible.
\end{acks}
\bibliographystyle{ACM-Reference-Format}


\appendix

\section{Implementation Details}

\subsection{Model Selection}
\label{app:model-selection}
Claude was selected for PASTA evaluation process based on three key strengths: (1) superior long-context handling, with models supporting context windows of up to 200K tokens, which is critical for processing large and fragmented legal or policy documents and aligning them with model card sections ~\cite{anthropic2023claude}; (2) consistent and schema-aligned output generation, as Claude performs well in producing structured compliance score tables and section-by-section explanations ~\cite{anthropic2023claude2}; and (3) seamless integration, as Claude is natively supported within LangChain, enabling efficient coordination across the multi-stage evaluation workflow.

\subsection{PASTA Model Card Section Adaptations}
\label{app:model-card-adaptations}

The PASTA model-card structure incorporates adaptations from existing model card templates to better support system-level documentation, regulatory alignment, and compliance evaluation. The specific section-level modifications are listed below:

\begin{itemize}
    \item \textbf{General Information}: Expanded to include detailed versioning practices and compliance-aware change tracking according to the documentation requirements in the EU AI Act~\cite{eu2024_regulation1689}.
    
    \item \textbf{Intended Uses}: Fully adopted from Mitchell et al.’s design of the Intended Uses section~\cite{Mitchell_2019}.
    
    \item \textbf{Existing Compliance Information}: Introduces explicit sections for existing terms and conditions as well as current legal compliance status to better support compliance evaluation.
    
    \item \textbf{System Data Information}: Combines the Evaluation Data and Training Data sections from Mitchell et al. and the Kaggle Model Card~\cite{Mitchell_2019, var0101_modelcards}. This unified structure is better suited for AI systems beyond machine learning models, where the distinction between training and evaluation datasets is often less meaningful. By consolidating these sections, we emphasize a system-level perspective on data provenance, collection practices, and bias mitigation, aligning more closely with the needs of compliance evaluation and regulatory transparency.
    
    \item \textbf{System Performance and Evaluation}: Expanded from the Quantitative Analyses section in the Kaggle Model Card~\cite{var0101_modelcards}. This section introduces disaggregated subgroup evaluations, detailed real-world testing contexts, adversarial testing scenarios, and robustness assessments. These additions provide a more granular view of system behavior across diverse conditions and user groups, which is critical for identifying potential compliance risks and ensuring fairness, safety, and reliability in real-world deployments.
    
    \item \textbf{Ethical Considerations}: Adopted from Mitchell et al.’s design of the ethical considerations section and structured based on the description in the Kaggle Model Card~\cite{Mitchell_2019, var0101_modelcards}.
    
    \item \textbf{Maintenance and Monitoring}: Includes detailed procedures for human oversight, real-time intervention, and systematic update tracking required by the EU AI Act~\cite{eu2024_regulation1689}.
\end{itemize}

\pagebreak

\subsection{Sample Evaluation Result}
\begin{table}[H]
\caption{Sample evaluation results represented in JSON format in a markdown table structure. Each policy clause is paired with a violation score and explanation, demonstrating how granular outputs are generated before aggregation into summaries.}
\centering
\small
\begin{tabular}{|l|p{7cm}|}
\hline
\textbf{Article} & \textbf{Evaluation} \\
\hline
1 & \{ "score": 0, "description": null \} \\
\hline
2 & \{ "score": 0, "description": null \} \\
\hline
3 & \{ "score": 3, "description": "The system name `Crop Health Monitor` may imply real-time monitoring capabilities not supported by the documented system functionality." \} \\
\hline
4 & \{ "score": 0, "description": null \} \\
\hline
5 & \{ "score": 2, "description": "The term `yield forecasting` is mentioned in the system name, but no forecasting methodology is provided, which introduces minor ambiguity regarding accuracy and compliance transparency." \} \\
\hline
\end{tabular}
\label{app:Example JSON-style Compliance Evaluation Output for the System Name Section}
\end{table}

\section{Technical Evaluation and Pair-wise Comparison Scoring Rubrics}

\label{app:curated-articles}
\subsection{Curated Articles for Evaluation}
The following policy articles were selected for expert evaluation:
\begin{enumerate}
    \item \textbf{EU AI Act}
    \begin{enumerate}
        \item Art.\ 5
        \item Art.\ 6
        \item Art.\ 9
        \item Art.\ 10
        \item Art.\ 11
        \item Art.\ 13
        \item Art.\ 14
        \item Art.\ 15
        \item Art.\ 23
        \item Art.\ 52
    \end{enumerate}

    \item \textbf{California Consumer Privacy Act (CCPA)}
    \begin{enumerate}
        \item \S{}1798.100
        \item \S{}1798.105
        \item \S{}1798.110
        \item \S{}1798.115
        \item \S{}1798.120
        \item \S{}1798.121
        \item \S{}1798.130
        \item \S{}1798.135
    \end{enumerate}
\end{enumerate}

\subsection{Pair--wise Comparison Scoring Rubrics}
\begin{table}[t]
\centering
\small
\setlength{\tabcolsep}{4pt}

\begin{minipage}[t]{0.48\textwidth}
\centering
\caption{Relevance Score Rubric (0--5) during the irrelevancy mapping stage. Scores $\leq 1$ indicate negligible impact, while higher scores reflect increasing importance for compliance assessment across policy articles.}
\label{tab:CCScore}

\begin{tabular}{|l|p{0.78\linewidth}|}
\hline
\textbf{Score} & \textbf{Description} \\
\hline
0 & Not Needed: The section has no impact on compliance assessment. Removing it does not affect this model card’s ability to determine compliance to one or multiple policy paragraphs. \\
\hline
1 & Marginal Contribution: Provides minimal information, with little impact on determining compliance to one or multiple policy paragraphs if removed. \\
\hline
2 & Some Contribution: Offers indirect relevance but does not directly determine compliance, or content applies but jurisdiction is out of scope (e.g., region not applicable). \\
\hline
3 & Moderate Contribution: Provides partial but significant evidence that helps assess compliance. \\
\hline
4 & High Contribution: The section is important for compliance assessment and removing it would make evaluation difficult. \\
\hline
5 & Essential Contribution: The section is indispensable. Without it, determining compliance to one or multiple policy paragraphs would be impossible or highly unreliable. \\
\hline
\end{tabular}
\end{minipage}

\vspace{2em}

\hfill
\begin{minipage}[t]{0.48\textwidth}
\centering
\caption{Violation scoring rubric (0--5) used in PASTA’s pairwise evaluation. The scale distinguishes between full compliance, ambiguous coverage, and clear violations.}
\label{tab:ViolationScore}
\vspace{4pt}

\begin{tabular}{|l|p{0.58\linewidth}|}
\hline
\textbf{Score} & \textbf{Description} \\
\hline
0 -- No Violation & The section clearly fulfills its intended purpose and presents no policy violations. \\
\hline
1 -- Minor Ambiguity & The section generally meets the policy requirement but lacks minor clarifications that would strengthen compliance confidence. \\
\hline
2 -- Ambiguous & The section is vague or incomplete in a manner that could introduce uncertainty. \\
\hline
3 -- Possible Risk & The section may reasonably lead to a policy violation due to unclear, incomplete, or permissive content. \\
\hline
4 -- Probable Violation & There are strong indications that the section permits or suggests behavior likely to conflict with the policy requirement. \\
\hline
5 -- Clear Violation & The section explicitly describes actions or conditions that directly violate or contradict the policy article. \\
\hline
\end{tabular}
\end{minipage}

\end{table}

\onecolumn
\subsection{Technical Evaluation Plots}
\label{app:eval-material}

\begin{figure}[H]
    \centering
    \includegraphics[width=0.7\linewidth]{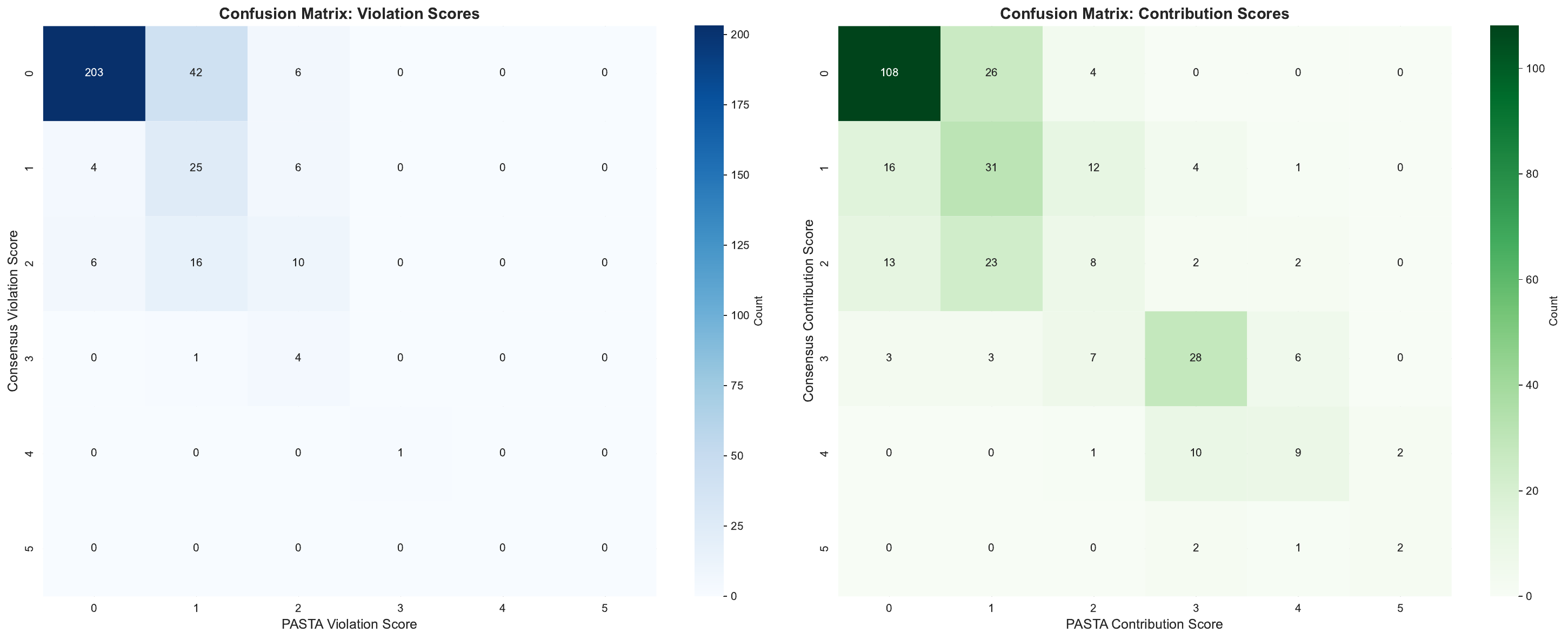}
    \caption{Confusion matrices comparing PASTA’s violation and contribution (relevance) scores against
expert consensus labels. Both matrices show strong diagonal structure, indicating high agreement on low- and mid-severity cases. Violation scores exhibit a conservative skew, with PASTA more frequently assigning lower scores than experts in higher-severity categories, resulting in density concentrated below the diagonal. Contribution score predictions display smoother alignment across the full 0--5 range, with broader coverage of higher categories and reduced compression relative to violation scores. These patterns suggest that PASTA maintains stable internal criteria, favoring caution in violation assessments while more closely reproducing expert distinctions in relevance judgments.}
\end{figure}

\begin{figure}[H]
    \centering
    \includegraphics[width=0.7\linewidth]{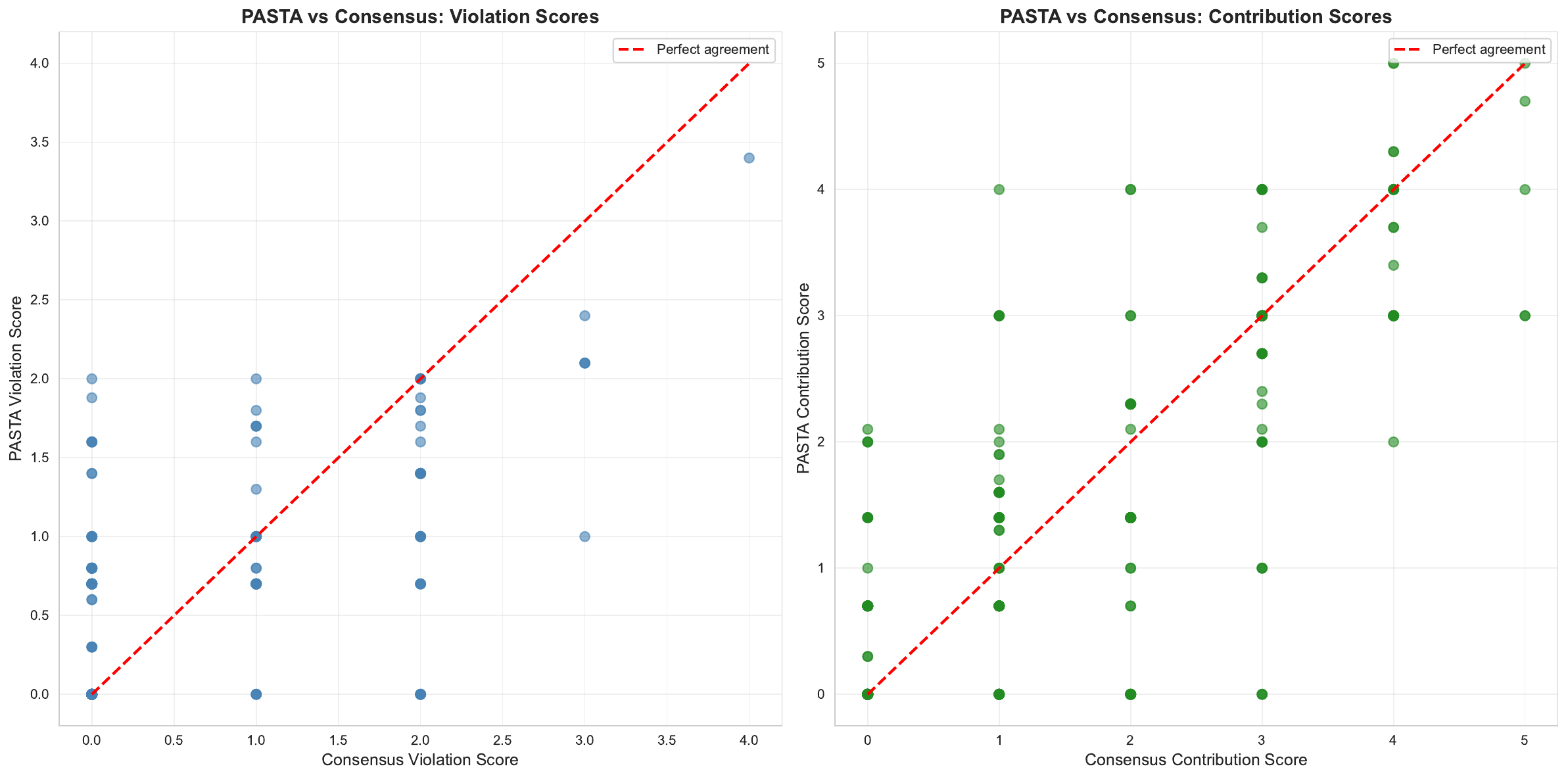}
    \caption{Scatter plots comparing PASTA’s violation and contribution scores to expert consensus, with a red dashed line indicating perfect agreement. Violation scores increasingly fall below the agreement line as expert-assigned severity rises, reflecting PASTA’s conservative tendency to underestimate high-risk cases when evidence for strong misalignment is ambiguous. In contrast, contribution scores span the full decision range and cluster tightly around the diagonal across all levels, demonstrating stronger calibration and finer-grained differentiation. Together, these trends illustrate that PASTA more reliably captures relevance relationships than high-severity violation judgments, while still exhibiting systematic—rather than stochastic—deviations from expert assessments.}
\end{figure}
\FloatBarrier

\onecolumn

\pagebreak
\subsection{PASTA UI}
\label{app:pasta-ui}
\begin{figure}[H]
    \centering
    \includegraphics[width=0.85\linewidth]{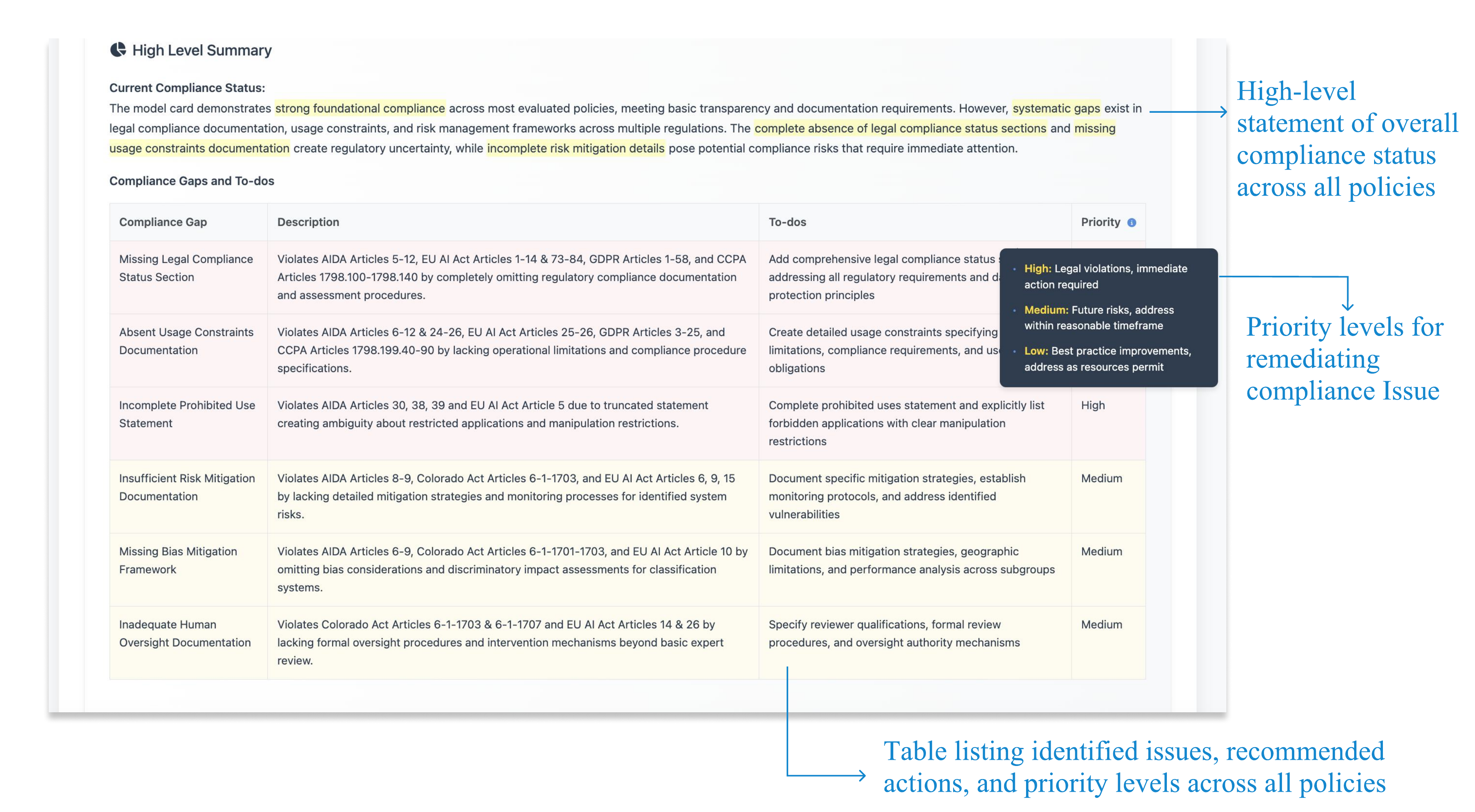}
    \caption{Interface view of an aggregated compliance summary showing an AI system’s overall status across multiple policies. The interface provides a high-level assessment alongside a table of identified issues, recommended actions, and their priority levels.}
    \label{fig:Example Overall Summary}
\end{figure}

\begin{figure}[H]
    \centering
    \includegraphics[width=0.85\linewidth]{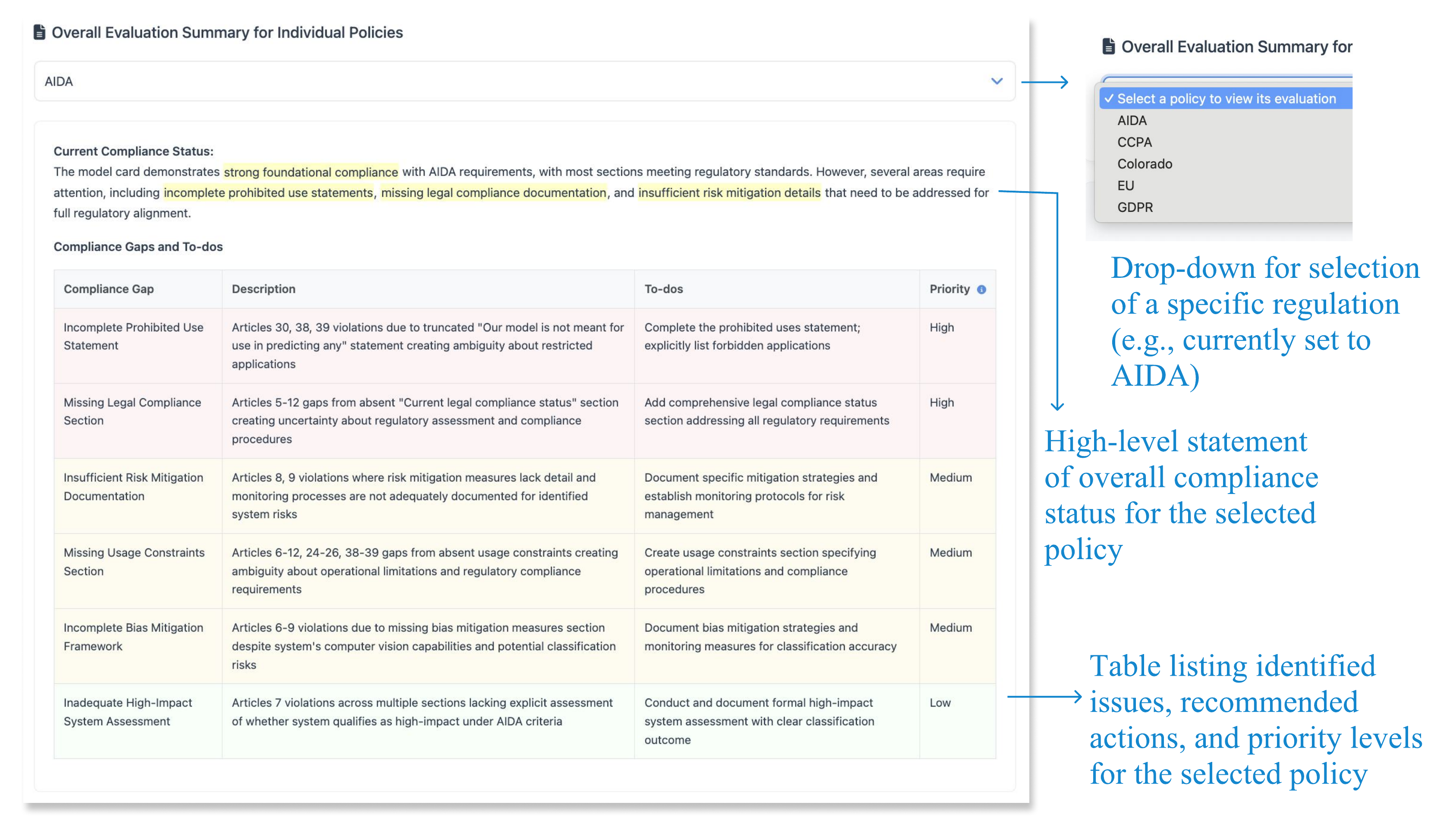}
    \caption{Interface view of compliance results filtered to a single policy (AIDA). The summary highlights key issues, recommended actions, and priorities, enabling users to explore compliance risks within one jurisdiction.}
    \label{fig:Example Policy-wise Summary for AIDA}
\end{figure}

\pagebreak
\FloatBarrier

\subsection{Model Card Template}
\label{app:model-card}

\small
\begin{longtable}{|P{2cm}|P{2cm}|P{13cm}|}
\caption{The customized model card structure used in PASTA, adapted from existing templates. It captures system-level information such as intended use, compliance status, data practices, and monitoring procedures in a standardized format for policy evaluation.} \\
\hline
\textbf{Category} & \textbf{Section} & \textbf{Description} \\
\hline
\endfirsthead

\hline
\textbf{Category} & \textbf{Section} & \textbf{Description} \\
\hline
\endhead

\multirow{5}{=}{General Information}
& System Name & State the full name of the AI system [<20 words]. \\
\cline{2-3}
& Versioning Information & Provide the date of the initial release, current version number, latest update date, and a brief explanation of your version tracking method (e.g., semantic versioning, Git tag). Clarify how major changes are documented or reviewed for compliance impacts. State N/A if not applicable. [40--60 words]. \\
\cline{2-3}
& Primary Developer/Org & List the individual or organization primarily responsible for the development and deployment of the system. If different entities manage operations, indicate that relationship. [20--30 words]. \\
\cline{2-3}
& Contact Information & Include an up-to-date email address and optionally a phone number or mailing address of the organization or individual accountable for the system. This contact should be able to respond to compliance inquiries or incident reports. [10--20 words]. \\
\cline{2-3}
& System Overview & Summarize the system's architecture, core components (e.g., model, APIs, databases), and technical functionality. Include whether the system is based on general-purpose AI models or tailored algorithms. Note if it operates autonomously or under human supervision. Specify development context (research prototype, internal tool, pilot, production), publication/open-source status, deployment environment(s), funding/sponsorship, and relevant jurisdictions/policies in scope. [40--60 words]. \\
\hline

\multirow{3}{=}{Intended Use}
& Primary Intended Uses & List the specific contexts, user needs, or operational environments for which the system was designed. Include the kinds of decisions, predictions, or outputs the system generates, and what user action (if any) it supports. [30--50 words]. \\
\cline{2-3}
& Primary Intended Users & Identify the primary audience or actors meant to operate, interact with, or rely on the system (e.g., clinicians, consumers, analysts). Clarify the level of expertise or training expected for responsible use. [25--35 words]. \\
\cline{2-3}
& Out-of-Scope Use Cases & Describe any uses, domains, or environments where the system is not intended to be applied. Include foreseeable misuse scenarios that may raise safety or fairness concerns, such as domains where the system is technically unreliable, lacks sufficient safeguards, or would create unacceptable risks. [30--50 words]. \\
\hline

\multirow{2}{=}{Existing Compliance Information}
& Terms and Conditions & Existing terms and conditions document: e.g. Usage Terms, Limitations of Liability, License Type, Third-party, Dependencies, Privacy and Data Handling, Compliance Acknowledgment. State N/A if not applicable. [25--40 words]. \\
\cline{2-3}
& Current legal compliance status & Indicate the system’s current status with respect to compliance with relevant AI regulations or frameworks if the system has been evaluated. Include results of any internal/external assessments, risk classification (e.g., high-risk under EU AI Act or impact assessment under AIDA), known gaps, or pending remediation steps. Cite assessment methods or tools used. State N/A if not applicable. [30--50 words]. \\
\hline

\multirow{4}{=}{System Data Information}
& Dataset Description & Describe each dataset used for your AI system if any. Include the purpose of the dataset, its size, source domains (e.g., medical records, social media), and whether the data represent diverse populations relevant to the intended use. Mention if the dataset includes personal or sensitive data and whether it was anonymized or synthesized. State N/A if not applicable. [40--60 words]. \\
\cline{2-3}
& Collection Method & Explain how each dataset was collected, including the data sources (e.g., surveys, web scraping), consent mechanisms, and whether third parties were involved. Clarify any limitations or assumptions made during collection and whether the data were collected directly from individuals or derived from secondary sources. State N/A if not applicable. [30--45 words]. \\
\cline{2-3}
& Bias Mitigation Measures & Detail the methods used to detect and reduce bias in data and model behavior. Include audits, rebalancing techniques, adversarial testing, subgroup performance analysis, and any manual interventions. Explain how outcomes were validated for fairness across different demographic groups. [30--45 words]. \\
\cline{2-3}
& Usage Constraints & Specify any limitations or conditions placed on system use, such as geographic, temporal, or technical constraints. Indicate legal restrictions (e.g., export controls), licensing terms (e.g., non-commercial only), or required approvals for deployment. Describe how inappropriate or unauthorized uses are prevented. State N/A if not applicable. \\
\hline

\multirow{4}{=}{System Performance and Evaluation}
& Summary of Performance & Summarize key evaluation metrics (e.g., accuracy, precision, recall, F1, fairness metrics) used to assess the system's performance. State N/A if not applicable. [30--45 words]. \\
\cline{2-3}
& Disaggregated Performance & Report system performance across different subgroups (e.g., age, gender, ethnicity, geography) relevant to the intended user population. Include metrics that highlight disparities or consistent trends. State how subgroup categories were selected, and whether performance gaps were identified and addressed. State N/A if not applicable. [30--50 words]. \\
\cline{2-3}
& Testing Contexts & Describe the environments and conditions under which the system was tested (e.g., simulated lab vs. real-world setting). Include geographical, social, and technical conditions, and explain how these contexts reflect actual deployment conditions. Note if testing included vulnerable or underrepresented populations. State N/A if not applicable. [30--50 words]. \\
\cline{2-3}
& Edge/Adversarial Testing & Detail evaluations conducted to assess the system's behavior under rare, extreme, or adversarial conditions. Include examples of inputs used, failure cases identified, and robustness strategies employed. Clarify any known limitations in edge performance and how these are mitigated. State N/A if not applicable. [30--45 words]. \\
\hline

\multirow{3}{=}{Ethical Considerations}
& Potential Risks and Harms & Identify risks the system may pose to health, safety, privacy, fundamental rights, or equity—especially in the context of foreseeable misuse. Include both technical and societal risks, noting affected groups (e.g., children, economically vulnerable populations). Highlight severity and likelihood. [30--45 words]. \\
\cline{2-3}
& Actions Taken & Summarize specific measures taken to prevent or mitigate the risks identified above. Include technical safeguards (e.g., red-teaming, input validation), process changes (e.g., oversight committees), or external interventions (e.g., third-party audits). Indicate the timeline and responsible parties for implementation. [30--50 words]. \\
\cline{2-3}
& Misuse Scenarios & Describe ways in which the system could be used improperly or outside its intended context, including foreseeable misuse that may result in safety risks, rights violations, or discriminatory outcomes. Indicate what technical or procedural safeguards are in place to prevent or detect such misuse. [30--50 words]. \\
\hline

\multirow{2}{=}{Maintenance and Monitoring}
& Human Oversight & Explain how human oversight is integrated into the system lifecycle. Describe the roles and responsibilities of human operators during development, deployment, and use. Indicate whether users can override system decisions, receive explanations, or intervene in real time, especially in high-risk or consequential scenarios. [30--45 words]. \\
\cline{2-3}
& Update Frequency & Indicate how often the AI system is updated (e.g., daily, monthly, irregularly) and describe the triggers or rationale for updates (e.g., model drift, new training data, regulatory changes). Explain the procedures in place to assess the impact of updates on safety, fairness, and compliance, and whether updates are logged and subject to review. [25--40 words]. \\
\hline

\end{longtable}
\FloatBarrier

\end{document}